\pdfoutput=1

\documentclass[11pt,a4paper]{article}

\usepackage[numbers, square, comma, sort&compress]{natbib}
\usepackage{amssymb}
\usepackage{amsthm}
\usepackage{amstext}
\usepackage{amsmath}
\usepackage{amsfonts}
\usepackage{verbatim}
\usepackage{geometry}
\usepackage{latexsym}
\usepackage{t1enc}
\usepackage{graphicx}
\usepackage[all]{xy}
\usepackage{makeidx}
\usepackage{slashed}
\usepackage{multicol}
\usepackage{alltt}

\allowdisplaybreaks

\newcommand \slsh [1] {\not\!{#1}}
\newcommand \e {\epsilon}

\newcommand{\secn}[1]{Section~\ref{#1}}

\def\beq{\begin{equation}}
\def\eeq{\end{equation}}
\def\beqa{\begin{eqnarray}}
\def\eeqa{\end{eqnarray}}
\def\eq#1{Eq.~(\ref{#1})}

\def\slash#1{#1 \hskip-0.45em /}

\def\spa#1.#2{\left\langle#1\,#2\right\rangle}
\def\spb#1.#2{\left[#1\,#2\right]}
\def\spash#1.#2{\spa{\smash{#1}}.{\smash{#2}}}
\def\spbsh#1.#2{\spb{\smash{#1}}.{\smash{#2}}}
\def\sand#1.#2.#3{%
  \left\langle\smash{#1^{-}}{\vphantom1}\right|{#2}%
  \left|\smash{#3^{-}}{\vphantom1}\right\rangle}
\def\sandp#1.#2.#3{%
  \left\langle\smash{#1^{-}}{\vphantom1}\right|{#2}%
  \left|\smash{#3^{+}}{\vphantom1}\right\rangle}
\def\sandpp#1.#2.#3{% 
  \left\langle\smash{#1^{+}}{\vphantom1}\right|{#2}%
  \left|\smash{#3^{+}}{\vphantom1}\right\rangle}
\def\sandpm#1.#2.#3{%
  \left\langle\smash{#1^{+}}{\vphantom1}\right|{#2}%
  \left|\smash{#3^{-}}{\vphantom1}\right\rangle}
\def\sandmp#1.#2.#3{%
  \left\langle\smash{#1^{-}}{\vphantom1}\right|{#2}%
  \left|\smash{#3^{+}}{\vphantom1}\right\rangle}

\def\ssand#1.#2.#3{%
  \left\langle\smash{#1}{\vphantom1}\right|{#2}%
  \left|\smash{#3}{\vphantom1}\right]}
\def\ssandp#1.#2.#3{%
  \left\langle\smash{#1}{\vphantom1}\right|{#2}%
  \left|\smash{#3}{\vphantom1}\right\rangle}
\def\ssandpp#1.#2.#3{%
  \left\langle\smash{#1}{\vphantom1}\right|{#2}%
  \left|\smash{#3}{\vphantom1}\right\rangle}

\def\proj{\flat}
\def\projdot#1.#2{k_{#1}^\proj\cdot k_{#2}^\proj}
\def\sandff#1.#2.#3{%
  \left\langle\smash{#1^{\proj,-}}{\vphantom1}\right|{#2}%
  \left|\smash{#3^{\proj,-}}{\vphantom1}\right\rangle}
\def\sandnf#1.#2.#3{%
  \left\langle\smash{#1^{-}}{\vphantom1}\right|{#2}%
  \left|\smash{#3^{\proj,-}}{\vphantom1}\right\rangle}
\def\sandfn#1.#2.#3{%
  \left\langle\smash{#1^{\proj,-}}{\vphantom1}\right|{#2}%
  \left|\smash{#3^{-}}{\vphantom1}\right\rangle}

\def\spa#1.#2{\left\langle#1\,#2\right\rangle}
\def\spb#1.#2{\left[#1\,#2\right]}

\numberwithin{equation}{section}

\begin{document}

\begin{titlepage}

\hbox{Edinburgh 2015/03}
\hbox{NIKHEF/2015-005}
\hbox{ITF-UU-15-02}

\vspace{25mm}

\begin{center}

   \Large{\sc{\bf A factorization approach to next-to-leading-power threshold logarithms}}

\end{center}

\vspace{8mm}

\begin{center}

D.~Bonocore$^1$, E.~Laenen$^{1,2,3}$, L.~Magnea$^4$, S.~Melville$^5$, 
L.~Vernazza$^6$, C.~D.~White$^5$ \\ [6mm]

\vspace{3mm}

\textit{$^1$Nikhef, Science Park 105, NL--1098 XG Amsterdam, The Netherlands} \\ 
\vspace{1mm}

\textit{$^2$ITFA, University of Amsterdam, Science Park 904, Amsterdam, 
The Netherlands} \\
\vspace{1mm}

\textit{$^3$ITF, Utrecht University, Leuvenlaan 4, Utrecht, The Netherlands} \\
\vspace{1mm}

\textit{$^4$Dipartimento di Fisica, Universit\`a di Torino and INFN, Sezione di Torino \\
Via P. Giuria 1, I-10125 Torino, Italy} \\
\vspace{1mm}

\textit{$^5$School of Physics and Astronomy, University of Glasgow, Glasgow 
G12 8QQ, UK} \\
\vspace{1mm}

\textit{$^6$Higgs Centre for Theoretical Physics, School of Physics and Astronomy, 
The University of Edinburgh, Edinburgh EH9 3JZ, Scotland, UK}

\end{center}

\vspace{8mm}

\begin{abstract}

\noindent Threshold logarithms become dominant in partonic cross
sections when the selected final state forces gluon radiation to be
soft or collinear. Such radiation factorizes at the level of
scattering amplitudes, and this leads to the resummation of threshold
logarithms which appear at leading power in the threshold variable.
In this paper, we consider the extension of this factorization to
include effects suppressed by a single power of the threshold
variable. Building upon the Low-Burnett-Kroll-Del Duca (LBKD) theorem,
we propose a decomposition of radiative amplitudes into universal
building blocks, which contain all effects ultimately responsible for
next-to-leading-power (NLP) threshold logarithms in hadronic cross
sections for electroweak annihilation processes. In particular, we
provide a NLO evaluation of the {\it radiative jet function},
responsible for the interference of next-to-soft and collinear effects
in these cross sections.  As a test, using our expression for the
amplitude, we reproduce all abelian-like NLP threshold logarithms in
the NNLO Drell-Yan cross section, including the interplay of real and
virtual emissions. Our results are a significant step towards
developing a generally applicable resummation formalism for NLP
threshold effects, and illustrate the breakdown of next-to-soft
theorems for gauge theory amplitudes at loop level.

\end{abstract}

\end{titlepage}

%%%%%%%%%%%%%%%%%%%%%%%%%%%%%%%%%%%%%%%%%%%%%%

\section{Introduction}
\label{intro}

It has long been known that the emission of soft and collinear gluons causes large 
corrections in perturbation theory, limiting its range of applicability. At the level of
partonic cross sections, these corrections take the form of {\it threshold logarithms} 
of ratios of physical momentum scales, which diverge as the total energy or 
transverse momentum of emitted gluons (for soft or collinear emission respectively)
becomes vanishingly small. Picking a dimensionless variable $\xi$ to measure the 
kinematic deviation from the threshold region, a differential cross section in this 
variable has a perturbative expansion of the form
\beq
  \frac{d \sigma}{d \xi} \, = \, \sum_{n = 0}^{\infty} \left( \frac{\alpha_s}{\pi} \right)^n \, 
  \sum_{m = 0}^{2 n - 1} \left[ c_{n m}^{(-1)}
  \left( \frac{\log^m \xi}{\xi} \right)_+ + \, c_{nm}^{(0)} \, \log^m \xi \, + \, \ldots \right] \, .
\label{thresholddef}
\eeq We refer to the leading contributions in this expansion,
characterized by the coefficients $c_{n m}^{(-1)}$, as {\it
  leading-power} (LP) threshold logarithms, while the $c_{n m}^{(0)}$
specify {\it next-to-leading-power} (NLP) threshold logarithms, and
the ellipsis denotes terms which are suppressed by additional powers
of $\xi$. A variety of approaches for describing LP threshold effects
exists in the literature, such as diagrammatic techniques based on
factorization
theorems~\cite{Sterman:1986aj,Catani:1989ne,Contopanagos:1997nh},
approximations using Wilson
lines~\cite{Korchemsky:1993uz,Korchemsky:1993xv}, renormalization
group arguments~\cite{Forte:2002ni}, dedicated effective field
theories~\cite{Becher:2006nr,Schwartz:2007ib,Bauer:2008dt,Chiu:2009mg}
and path integral techniques~\cite{Laenen:2008gt}. Crucial to all
these approaches is the notion of {\it factorization}, corresponding
to the intuitive idea that long-distance effects in perturbative cross
sections must have a universal nature, tied with the wave functions of
the scattering states, and it must be possible to disentangle them
from the contributions of short-distance high-energy exchanges.

At the level of scattering amplitudes, long-distance effects at leading power manifest
themselves in the form of infrared and collinear divergences, which are known
to factorize~\cite{Collins:1989bt,Dixon:2008gr}. For a generic $n$-point gauge theory
scattering amplitude, this factorization takes the schematic form
\beq
  {\cal A}_{(n)} \, = \, \prod_{i = 1}^n \left[ \frac{J_i}{{\cal J}_i} \right] \cdot {\cal S}_{(n)} 
  \cdot {\cal H}_{(n)} \, .
\label{factorised}
\eeq
Here the jet functions $J_i$ (one for each external parton) contain universal collinear 
singularities which depend only on the color and spin quantum numbers of the 
external states, the soft function ${\cal S}_{(n)}$ collects all soft singularities, which correlate
all partons but do not depend on their energies and spins, and ${\cal H}_{(n)}$ is a hard 
function, which is process dependent and infrared finite. One must finally divide each 
jet by its eikonal counterpart ${\cal J}_i$, to correct for the fact that soft and collinear 
divergences have been double-counted. The factorization in \eq{factorised} leads to
exponentiation of soft and collinear factors in terms of a restricted set of anomalous 
dimensions: color singlet anomalous dimensions are known to three loops, since they
can be extracted from the poles of partonic form factors~\cite{Moch:2005id,Moch:2005tm}; 
the soft anomalous dimension matrix, on the other hand, is known at two loops for both 
massless and massive partons~\cite{Aybat:2006wq,Aybat:2006mz,Becher:2009cu,
Gardi:2009qi,Becher:2009qa,Kidonakis:2009ev,Becher:2009kw,Ferroglia:2009ii,
Ferroglia:2009ep,Mitov:2009sv}; the development of the necessary techniques to 
extend these results to the three-loop order and beyond is under way~\cite{Mitov:2010rp,
Gardi:2010rn,Gardi:2011wa,Gardi:2011yz,Correa:2012nk,Henn:2012qz,Dukes:2013wa,
Henn:2013wfa,Gardi:2013ita,Dukes:2013gea,Gardi:2013saa,Grozin:2014axa,
Falcioni:2014pka,Grozin:2014hna}. 

The factorization of leading-power infrared enhancements at amplitude
level, embodied by \eq{factorised}, ultimately leads to the
resummation of LP threshold logarithms for a large class of
infrared-safe observables, under mild assumptions concerning the
behavior of the associated real radiation~\cite{Laenen:2004pm}. In
most cases, current knowledge of the anomalous dimensions allows one to
perform this resummation up to N$^2$LL or approximate N$^3$LL
accuracy. Given this increasing theoretical precision, and given the
growing demands of current collider experiments, it is natural to
attempt to extend existing results to the next set of
contributions\footnote{Recall that another important set of terms can
  be controlled to all orders, and can be shown to exponentiate in
  some cases: these are the contributions that have support only on
  the threshold, proportional to $\delta(\xi)$ if $\xi$ is the
  threshold variable~\cite{Parisi:1980xd, Eynck:2003fn}. These terms
  will not be discussed here.} in \eq{thresholddef}: indeed, these
effects are known to be potentially significant~\cite{Kramer:1996iq,
  Harlander:2001is,Harlander:2002wh,Catani:2001ic,Catani:2003zt}, as
might be expected from the fact that they are still singular as $\xi
\rightarrow 0$, albeit integrably so.  Furthermore, there are
theoretical reasons to expect that it should be possible to organize
NLP threshold effects. Indeed, it has been known for a long time that
soft gluon radiation effects at NLP can still be expressed in a
universal way in terms of the non-radiative matrix element: this is
the content of the Low-Burnett-Kroll
theorem~\cite{Low:1958sn,Burnett:1967km}. This theorem was extended to
the case of massless theories, where collinear effects become
important, by Del Duca~\cite{DelDuca:1990gz}.

With these motivations, in recent years several steps were taken towards a deeper 
understanding of NLP threshold effects and ultimately towards a resummation of NLP 
threshold logarithms~\cite{Dokshitzer:2005bf,Laenen:2008gt,Laenen:2008ux,
Moch:2009mu,Grunberg:2009yi,Moch:2009hr,Soar:2009yh,Laenen:2010uz,
Almasy:2010wn,Ball:2013bra,Altinoluk:2014oxa,Apolinario:2014csa,deFlorian:2014vta,
Presti:2014lqa,Bonocore:2014wua}, including results for a complete resummation of 
certain towers of NLP logs in specific processes. A fully general resummation prescription, 
applicable to all processes for which threshold logarithms can be resummed, has however 
yet to be developed. In particular, one must carefully disentangle the interplay between 
the soft expansion, which is well under control, and hard collinear effects, which can 
also be the source of threshold logarithms.

In this paper, we will build on earlier results of
Refs.~\cite{Laenen:2008gt,Laenen:2010uz, Bonocore:2014wua}. The first
of these papers used path integral methods to perform a systematic
expansion in powers of the momentum of emitted gluons, thus deriving a
generalization of the soft-collinear factorization formula of
\eq{factorised}, valid up to next-to-soft order. The second paper
rederived these results using a diagrammatic approach, and also
performed a first check of the formalism, applying it to the case of
Drell-Yan production of a color-singlet vector boson. It was shown
that the next-to-soft Feynman rules correctly reproduce the known NLP
logarithms in the double-real-emission contribution to the K factor
for this process, up to NNLO~\cite{Matsuura:1988nd,
  Matsuura:1989sm,Matsuura:1988wt,Hamberg:1991np,Matsuura:1991pc}. Whilst
Refs.~\cite{Laenen:2008gt,Laenen:2010uz} addressed the factorization
and exponentiation properties of next-to-soft effects, the results
were limited, as is the earlier work of
Refs.~\cite{Low:1958sn,Burnett:1967km}, by the fact that collinear
singularities were not properly accounted for.  In the case of
Drell-Yan production, final state gluons are forced to be (next-to-)
soft, so the interplay of soft and collinear enhancements arises only
due to the presence of virtual hard-collinear gluons, which distort
the emission spectrum of soft gluons at next-to-leading power in the
soft expansion. In essence, in the context of \eq{factorised}, one
must also include the possibility of soft emissions from the jet
functions which contain all collinear radiation.

The tools to tackle this problem, and thus to extend the Low-Burnett-Kroll theorem 
\cite{Low:1958sn,Burnett:1967km} to collinear-singular amplitudes, were developed 
by Del Duca~\cite{DelDuca:1990gz}, who proposed a generalization of the standard 
hard-soft factorization, designed to organize the effects of next-to-soft radiation 
in the massless limit. A further step forward was recently taken by some of us
in~\cite{Bonocore:2014wua}, where NLP threshold corrections to NNLO Drell-Yan 
production were precisely classified according to their soft or collinear origin, using 
the {\it method of regions} developed in Refs.~\cite{Beneke:1997zp,Pak:2010pt,
Jantzen:2011nz}\footnote{The method of regions has also been recently applied,
in a slightly different context, to characterise threshold effects in Higgs production 
via gluon fusion up to N$^3$LO, in Refs.~\cite{Anastasiou:2013mca,Herzog:2014wja},}. 
The results of Ref.~\cite{Bonocore:2014wua} imply that in the case of electroweak 
annihilation cross sections the configurations needed to bridge the gap between the
soft expansion and the threshold expansion are precisely the collinear-enhanced ones
which are taken into account in Ref.~\cite{DelDuca:1990gz}.

Our aim in the present paper is to build on the results of
Refs.~\cite{DelDuca:1990gz,
  Laenen:2008gt,Laenen:2010uz,Bonocore:2014wua}, in the light of
\eq{factorised}, to construct a precise formulation accounting for all
NLP threshold logarithms for electroweak annihilation cross
sections. With this in mind, we will improve upon the approach
of~\cite{DelDuca:1990gz} by taking into account the factorization of
soft modes (which were included in the hard interaction in
Ref.~\cite{DelDuca:1990gz}), by explicitly taking care of the double
counting of soft-collinear singularities, and by discussing the role
played by the factorization vectors in the factorization of collinear
from soft modes. This will lead us to a general expression for the
radiative scattering amplitude, similar to that derived
in~\cite{DelDuca:1990gz}, which contains all universal ingredients
needed to organize NLP threshold logarithms in the process at hand.
We will then test our formalism, once again in the context of
Drell-Yan production, by reproducing the known NLP logarithms in the
mixed real-virtual contribution to the K factor for this process at
NNLO, focussing for simplicity on the abelian-like term proportional
to $C_F^2$. We will describe how our formalism can be generalized to
include purely non-abelian contributions, but we leave more detailed
phenomenological analysis to future work.

In parallel to the above developments, a recent body of work has
examined the next-to-soft behaviour of gauge and gravity scattering
amplitudes from a more formal point of
view~\cite{Cachazo:2013hca,Cachazo:2013iea,Strominger:2013jfa,He:2014laa,
  Cachazo:2014fwa,Casali:2014xpa,Schwab:2014xua,Bern:2014oka,He:2014bga,
  Larkoski:2014hta,Cachazo:2014dia,Afkhami-Jeddi:2014fia,Adamo:2014yya,
  Bianchi:2014gla,Bern:2014vva,Broedel:2014fsa,He:2014cra,Zlotnikov:2014sva,
  Kalousios:2014uva,Du:2014eca,Luo:2014wea}, leading to a general
theorem for the structure of next-to-soft corrections at tree-level
and beyond. Particularly relevant in this regard is the study of
Ref.~\cite{Larkoski:2014bxa}, which approaches the classification of
next-to-soft emissions from the viewpoint of soft-collinear effective
theory (SCET), including also a detailed discussion of collinear
effects. The relationship between this body of work and the approach
of this paper has been recently clarified in Ref.~\cite{White:2014qia}
(see also~\cite{White:2011yy}). Here we will follow up on this, by
pointing out the implications of our results for loop corrections to
next-to-soft theorems: our results broadly agree with, and
substantiate with a concrete example, the arguments of
Refs.~\cite{Bern:2014vva,Larkoski:2014bxa}, showing that in the
presence of collinear singularities next-to-soft theorems for
scattering amplitudes require non-trivial corrections at one loop and
beyond.

The structure of our paper is as follows. In \secn{sec:review} we
review in detail the modified factorization formula for next-to-soft
emissions in the presence of collinear singularities proposed in
Ref.~\cite{DelDuca:1990gz}, which we refine in several respects to
emphasize the role of soft factors. As in Ref.~\cite{DelDuca:1990gz},
our final expression for the radiative amplitude at next-to-soft level
contains a universal function describing the emission of a soft gluon
from inside a jet, which we call the {\it radiative jet function}. We
calculate this function explicitly at one loop in
\secn{sec:jetemit}. In \secn{sec:DY}, we assemble all necessary
ingredients in order to reproduce the NLP threshold contributions to
the Drell-Yan $K$-factor, providing a strong validation of our
expression for the radiative amplitude. We discuss our results in
\secn{sec:conclusion} before concluding. Technical details are
presented in an Appendix.

%%%%%%%%%%%%%%%%%%%%%%%%%%%%%%%%%%%%%%%%%%%%%%

\section{Organizing NLP threshold logarithms}
\label{sec:review}

In this section, we discuss in detail the construction of an
expression, at the amplitude level, that directly leads to organizing
threshold logarithms up to next-to-leading power. As already mentioned
in the Introduction, a class of NLP threshold corrections arising from
soft gluons at next-to-leading power in the soft expansion was
analyzed in Refs.~\cite{Laenen:2008gt, Laenen:2010uz}. The resulting
factorization can describe those processes in which all threshold
contributions arise only from soft emissions, but it will fail when
collinear singularities are present in addition. This problem is
shared by earlier works on next-to-soft corrections, such as the
well-known investigation by Low~\cite{Low:1958sn}, who considered the
emission of soft photons from hard scalar emitting particles, and
showed how the resulting matrix element can be expressed in terms of
(derivatives of) the non-radiative amplitude\footnote{See, for
  example, Appendix C of Ref.~\cite{Laenen:2008gt} for a modern review
  of this material.}. This seminal theorem was generalized to
fermionic emitters in Ref.~\cite{Burnett:1967km}, and the combined
result is known as the Low-Burnett-Kroll theorem. This is still not
the full story, however, as the results of
Refs.~\cite{Low:1958sn,Burnett:1967km} apply only to particles with a
non-vanishing mass $m$, so that one may take the energy $E$ of soft
gluons to zero in the parametrically well-defined limit $E/m
\rightarrow 0$. One must clearly amend this argument for massless
external particles, which in practice involves carefully disentangling
collinear singularities (associated with $m\rightarrow 0$) from those
associated with the soft expansion. This important generalization was
achieved by Del Duca~\cite{DelDuca:1990gz}, who derived an expression
for the radiative amplitude valid in the extended region $m^2/Q \leq E
< m$, where $Q$ is the energy scale associated with the hard
interaction. Taking $m/Q$ parametrically to zero, one obtains a
modification of \eq{factorised}, which is correct (at least in
principle) up to next-to-soft level. For the purposes of studying
threshold corrections, all these results are necessary, and in this
paper we refer to this body of knowledge as the Low-Burnett-Kroll-Del
Duca (LBKD) theorem\footnote{The classification of NLP threshold
  effects from collinear regions has also been extensively discussed
  using effective field theory methods in the recent
  study~\cite{Larkoski:2014bxa}.}.

Let us begin by reviewing the results of Ref.~\cite{DelDuca:1990gz}, and placing them 
in the context of \eq{factorised}, which was developed in subsequent years. This will 
enable us to refine the original treatment, in particular explicitly extracting soft effects, 
studying the dependence on reference vectors defining the collinear region, and 
addressing the double counting of contributions which are both soft and collinear. 
We start our analysis by examining in more detail the soft-collinear factorization 
formula of \eq{factorised} for the specific case of two hard colored particles, 
which we may write as
\beqa
\label{softcolfac}
  {\cal A} \left( \frac{Q^2}{\mu^2}, \alpha_s (\mu^2), \epsilon \right) & = & 
  {\cal H} \Big( \{ p_i \}, \{ n_i \}, \alpha_s (\mu^2), \epsilon \Big)
  \times {\cal S} \Big( \{ \beta_i \}, \alpha_s (\mu^2), \epsilon \Big) \nonumber \\ 
  & & \qquad \quad \times \, \prod_{i = 1}^2 \left[ \frac{J_i \left( p_i, n_i, \alpha_s (\mu^2), 
  \epsilon \right)}{{\cal J}_i \left( \beta_i, n_i, \alpha_s (\mu^2), \epsilon \right)} \right] \, ,
\eeqa
Here we work in $d = 4 - 2 \epsilon$ dimensions, and $p_i$ ($\beta_i$) is the
four-momentum (four-velocity) of the $i$-th hard particle. Again, $Q$ is the energy 
scale associated with the hard interaction: for definiteness, we take here  $Q^2 = 
(p_1 + p_2)^2 > 0$. 

The partonic jet functions appearing in \eq{softcolfac} are defined by~\cite{Dixon:2008gr}
\beq
  J \Big( p, n, \alpha_s(\mu^2), \epsilon \Big) u(p) \, = \, \left\langle 0 \left| \Phi_n (\infty, 0)
  \psi(0) \right| p \right\rangle \, ,
\label{Jdef}
\eeq
where $\psi(x)$ is a quantum field inserted to absorb the external incoming parton 
with momentum $p$. The factor $\Phi_n$, on the other hand, represents a Wilson 
line stretching from the absorption point to infinity, along a direction fixed by an 
auxiliary vector $n^\mu$, according to the definition
\beq
  \Phi_n ( \lambda_2, \lambda_1) \, = \, {\cal P} \exp \left[ i g_s
  \int_{\lambda_1}^{\lambda_2} d \lambda\, n \cdot A(\lambda n) \right] \, .
\label{phidef}
\eeq
The presence of the Wilson line ensures that the definition is gauge-invariant. As was 
remarked in the Introduction, one must further introduce in \eq{softcolfac} the eikonal 
counterpart of the jet functions, ${\cal J}_i$, in order to avoid the double counting of 
soft-collinear contributions. The eikonal jet is defined as
\beq
  {\cal J} \Big( \beta, n,\alpha_s(\mu^2), \epsilon \Big) \, = \, \left\langle 0 \left|
  \Phi_n (\infty, 0) \Phi_{\beta} (0, - \infty) \right| 0 \right\rangle \, .
\label{eikJdef}
\eeq
Note that Eqs. (\ref{Jdef}) and (\ref{eikJdef}) apply to incoming particles: for outgoing 
particles, one would have to reverse the direction of the Wilson lines.

A key point of this discussion is the fact that one must associate with each leg of 
momentum $p_i$ an auxiliary four-vector $n_i$. This ensures gauge invariance, but,
more interestingly, it can be physically interpreted as providing a means for measuring 
collinearity with respect to $p_i$. The $n_i$'s are `factorization vectors', and the full 
amplitude cannot depend on them, much as it cannot depend on the factorization and 
renormalization scale $\mu$. In fact, singular dependence on $n_i$ cancels between 
each jet and its eikonal counterpart, while non-singular $n_i$-dependent terms cancel 
between the jets and the hard function.

In order not to introduce in the jet functions $J_i$ spurious collinear singularities
not associated with emissions from the $i$-th hard parton, it is customary in factorization
studies~\cite{Dixon:2008gr} to take $n_i^2 \neq 0$. This has the advantage of allowing 
a cleaner identification of each subset of singular contributions, but it has the drawback 
of introducing a more complicated functional dependence on the factorization vectors. 
In contrast, a more physical viewpoint is to consider the $n_i$ vectors as standing in 
to replace the other hard partons in the process: this is what is typically done in 
effective field theory calculations~\cite{Larkoski:2014bxa}, and was also the approach 
followed in Ref.~\cite{Bonocore:2014wua}. This choice leads to much simpler expressions, 
at the price of a degree of ambiguity in the classification of singular regions. Here, we 
will follow Ref.~\cite{DelDuca:1990gz}, and define two dimensionless light-like vectors 
$\hat{n}_i$ in directions opposite to $p_i$, and such that
\beq
  \hat{n}_i^2 \, = \, 0, \qquad  \qquad \hat{n}_i \cdot p_i \, = \, Q \, .
\label{nihatdef}
\eeq
In the calculations of Sections~\ref{sec:jetemit} and~\ref{sec:DY}, it will in fact be more
convenient to work with the dimensionful vectors
\beq
  n_i \, = \, \frac{Q}{2} \, \hat{n}_i \, , \qquad \qquad n_i \cdot p \, = \, \frac{Q^2}{2} \, ;
\label{nidef}
\eeq
note that for a two-parton process, such as that of \eq{softcolfac}, the definitions of 
\eq{nidef} can be implemented simply by choosing
\beq
  n_1 \, = \, p_2 \, ,\qquad  \qquad n_2 \, = \, p_1 \, ,
\label{n1p2}
\eeq
an identification that will be useful in \secn{sec:DY}. 

Let us now turn to the soft function, which collects infrared singularities associated 
with the leading (eikonal) term in the momentum expansion of emitted gluons. For 
two-parton annihilation, it is defined by
\beq
  {\cal S} \left( \beta_1 \cdot \beta_2, \alpha_s(\mu^2), \epsilon \right) \, = \, 
  \left\langle 0 \left| \Phi_{\beta_2} (\infty,0) \Phi_{\beta_1} (0, - \infty) 
  \right| 0 \right\rangle \, .
\label{Sdef}
\eeq
These expressions for the soft and jet functions implicitly define also the 
process-dependent hard function: for a given process, one defines ${\cal H}$ 
via \eq{softcolfac}, by taking all perturbative contributions to the soft and jet 
functions over to the left-hand side. The hard function then depends on the
auxiliary vectors $n_i$ precisely in such a way as to cancel the finite parts 
of their contribution to the jet functions, so that the physical amplitude is
independent of each $n_i$. 

As assembled in \eq{softcolfac}, the double counting of soft-collinear regions
is solved by assigning soft-collinear poles to the soft function, and constructing
ratios of jets and eikonal jets which contain only hard collinear singularities. An
alternative arrangement of the same expression is to assign the soft-collinear
poles to the jets, and construct the ratio of the soft function and the eikonal jets,
which contains only soft wide-angle radiation. This ratio defines the {\it reduced 
soft function},
\beq
  \overline{\cal S} \left( \frac{\beta_1 \cdot \beta_2}{\beta_1 \cdot n_1 \, \beta_2
  \cdot n_2}, \alpha_s(\mu^2), \epsilon \right) \, = \, 
  \frac{{\cal S} (\beta_1 \cdot \beta_2, \alpha_s(\mu^2), \epsilon)}{\prod_i 
  {\cal J}_i (\beta_i \cdot n_i, \alpha_s(\mu^2), \epsilon)} \, .
\label{sbardef}
\eeq
On the left-hand side, we have indicated that the arguments of the soft and jet 
functions must combine to form a ratio constructed to be invariant under the
rescalings $\beta_i \rightarrow \lambda _i \beta_i$. As discussed in detail 
in~\cite{Dixon:2008gr,Gardi:2009qi,Gardi:2009zv}, for light-like $\beta_i$ this 
invariance is broken for the soft function alone, as well as for the eikonal jets, due
to the presence of collinear divergences in either factor. When the individual 
factors are combined into the reduced soft function, as in \eq{sbardef}, collinear 
poles cancel and the invariance is restored. If, on the other hand, we work with
light-like $n_i$, the spurious collinear divergences associated with the Wilson 
lines in the $n_i$ directions do not cancel in $\overline{\cal S}$, so the expected
invariance under the rescalings $n_i \rightarrow \kappa_i n_i$ is not restored, 
as seen from the argument in \eq{sbardef}.

\begin{figure}
\begin{center}
  \scalebox{0.8}{\includegraphics{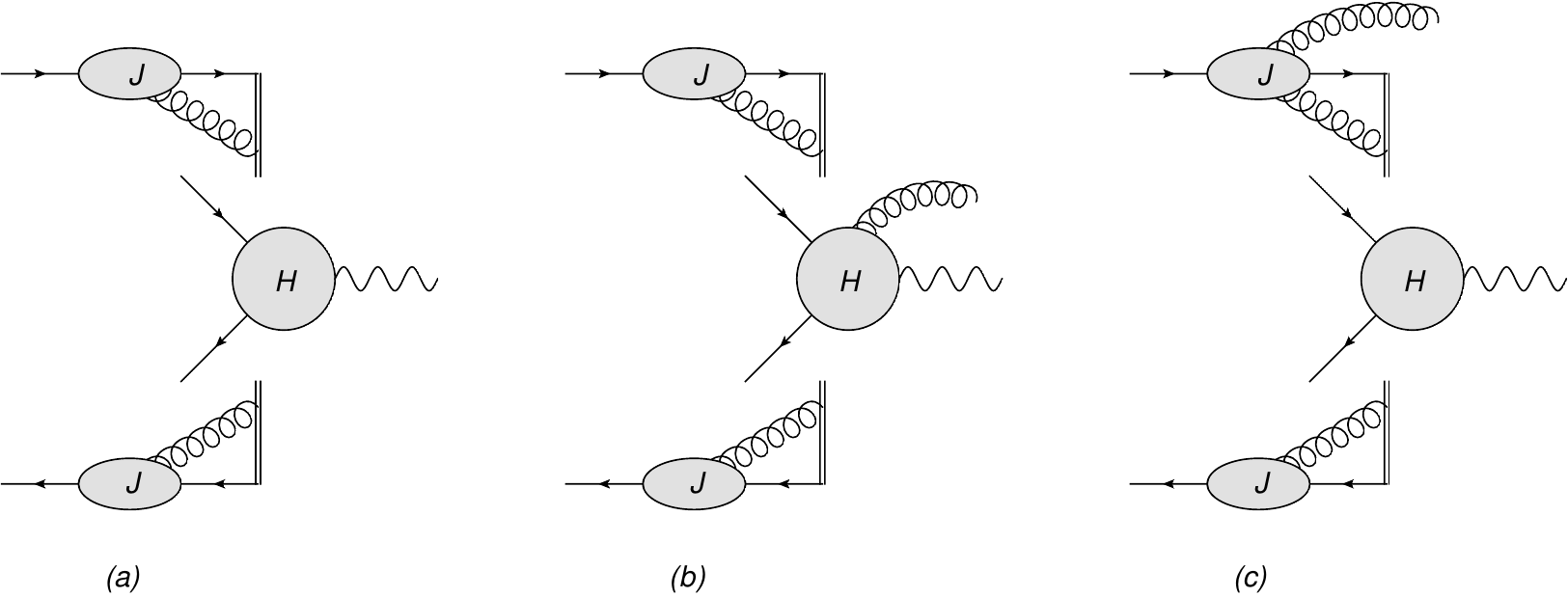}}
  \caption{Schematic depiction of the factorization of the amplitude into 
  the non-collinear function $H$ of \eq{Hdef} and external jet functions: (a) portrays
  the non-radiative amplitude, while (b) and (c) contribute to the radiation of an 
  extra gluon.}
\label{fig:HJfac}
\end{center}
\end{figure}

Making use of \eq{sbardef}, we may now rewrite schematically the amplitude in 
\eq{softcolfac} as
\beq
  {\cal A} \, = \, {\cal H} \times \bar{\cal S} \times \prod_{i = 1}^2 J_i \, ,
\label{ampdefsbar}
\eeq
where the functions $\{J_i\}$ contain all relevant information associated with the 
collinear regions. Furthermore, in the remainder of this section, we will follow 
Ref.~\cite{DelDuca:1990gz} and define a `non-collinear' factor
\beq
  H \, \equiv \, {\cal H} \times \bar{\cal S} \, ,
\label{Hdef}
\eeq
where the reduced soft function is absorbed into the hard function. The factorized 
structure of the amplitude is then as shown in Fig.~\ref{fig:HJfac}(a).
Let us now describe how to generalise \eq{softcolfac} to NLP level, building on 
Ref.~\cite{DelDuca:1990gz}. First of all we wish to isolate the contributions to the 
radiative amplitude where the extra gluon is emitted by a collinearly enhanced 
configuration. With this in mind, and denoting the amplitude with an additional
gluon emission by ${\cal A}_\mu$, one may naturally write
\beq
  {\cal A}_\mu \, \epsilon^\mu (k) \, = \, {\cal A}_\mu^{J} \, \epsilon^\mu (k)
  + {\cal A}_\mu^H \, \epsilon^\mu (k) \, ,
\label{Amudef}
\eeq
where we are suppressing color indices, $\epsilon_\mu (k)$ is the polarization vector 
of the extra gluon, and ${\cal A}_\mu^J$ (${\cal A}_\mu^H$) represent emissions from 
the jet (hard) functions, respectively. The amplitude for emission from collinear 
configurations can be defined as
\beq
  {\cal A}_\mu^J \, = \, \sum_{i = 1}^2 
  H(p_i - k; p_j, n_j) \, J_\mu (p_i, k, n_i) \, \prod_{j \neq i} J (p_j, n_j) \, \equiv \,
  \sum_{i = 1}^2 {\cal A}_\mu^{J_i} \, .
\label{AmuJ}
\eeq
Here, for brevity, we have not displayed the dependence on the coupling and on 
$\epsilon$; we have introduced in $H$ the notation of separating with a semi-colon
the `active' momentum (here the momentum of the radiating leg) from the other 
vectors appearing in the function; most importantly, we have introduced the 
{\it radiative jet function},
\beq
  J_\mu \left( p, n, k, \alpha_s(\mu^2), \epsilon \right) u(p) \, = \, 
  \int d^d y \,\, {\rm e}^{ - {\rm i} (p - k) \cdot y} \, \left\langle 0 \left| \,
   \, \Phi_{n} (y, \infty) \, \psi (y) \, j_\mu (0) \, \right| p \right\rangle \, ,
\label{Jmudef}
\eeq
representing the emission of a gluon from inside a jet function. It is useful to compare 
this definition with the non-radiative jet function defined in \eq{Jdef}. Here the incoming 
particle (a quark of momentum $p$ in this case) is absorbed by the field at position $y$, 
but, along the way, the insertion of the chromo-electric current $j_\mu$ causes the 
loss of momentum $k$ through the emission of a gluon. The gauge phase is then 
translated to infinite distance by the customary Wilson line. Finally the fact that fields 
are evaluated at different space-time points requires a Fourier transform in order to 
get the momentum-space correlator. For the purposes of the present paper we can 
take for the current the QED-like expression
\beq
  j^\mu_a (x) \, = \, \overline{\psi} (x) \, \gamma^\mu \, T_a \, \psi (x) \, ,
\label{abcurr}
\eeq where $T^a$ is a color generator in the fundamental
representation. In fact, we will continue to suppress color indices
and matrices throughout the paper, except when especially relevant,
since we will explicitly consider only QED-like diagrams and focus on
contributions to the cross section proportional to
$C_F^2$. Ultimately, however, we will need to generalize the
definition in \eq{Jmudef} to the full non-abelian theory, and to
gluon-originated jets as well. A natural generalization of \eq{abcurr}
is to consider the conserved (though not {\it covariantly} conserved)
non-abelian current (see e.g. Ref.~\cite{Weinberg:1996kr}) 
\beq {\mathfrak j}^\mu_a (x) \, = \,
f_a^{\phantom{a} b c} \, F^{\mu \nu}_b (x) \, A_{\nu c} (x) +
\overline{\psi} (x) \, \gamma^\mu \, T_a \, \psi (x) \, ,
\label{nabcurr}
\eeq 
where $f_a^{\phantom{a} b c}$ are $SU(N)$ structure
constants. This choice has all the physical characteristics required
for our definition: for example it allows radiation from a gluon jet,
and it includes the proper non-abelian corrections to gluon radiation
from a quark jet. There are, however, significant differences between
the two cases concerning the application of relevant Ward identities
and the renormalization properties, which modify to some extent the
reasoning given below in the `abelian' case, without disrupting the
general structure of the argument; therefore, we leave the full
treatment of the non-abelian radiative jet to future work. We note
that in either case \eq{Jmudef} defines a universal object, depending
solely on the properties of the given external particle (including its
spin), but not on the details of the specific hard interaction
process. We will explore further how to calculate this object in
\secn{sec:jetemit}; here we begin by noting that the radiative jet
function defined in \eq{Jmudef}, with the current in \eq{abcurr},
obeys the simple Ward identity~\cite{DelDuca:1990gz} \beq k^\mu \,
J_\mu \left(p, n, k, \alpha_s(\mu^2), \epsilon \right) \, = \, q \, J
\left( p, n, \alpha_s(\mu^2), \epsilon \right) \, ,
\label{WardJ}
\eeq
where $q = \pm 1$ according to whether the momentum $p$ is incoming or outgoing 
respectively ($q$ would represent the electric charge in a QED calculation). We may 
similarly consider the Ward identity for the entire amplitude: in this case, it takes the 
form
\beq
  k^\mu \, {\cal A}_\mu \, = \, 0 \, ,
\label{Wardamp}
\eeq
which in turn implies
\beq
  k^\mu \, {\cal A}^H_\mu \, = \, - \, k^\mu \, {\cal A}^J_\mu \, .
\label{Ward}
\eeq
We can use this result to relate emissions from inside the hard function, collected in
the function ${\cal A}^H_\mu$, to those from the jet functions, represented by ${\cal 
A}^J_\mu$. Indeed, note that so far ${\cal A}^H_\mu$ has been defined as a matching
condition: the Ward identity shows that it is not independent from ${\cal A}^J_\mu$.
To be explicit, we may use Eqs.~(\ref{AmuJ}) and~(\ref{WardJ}) to find that
\beqa
\label{AHexpand}
  k^\mu {\cal A}^J_\mu (p_i, k) & = & \sum_{i = 1}^2 q_i \, H (p_i - k; p_j, n_j) 
  \prod_{j = 1}^2 J (p_j, n_j) \\ 
  & = & \sum_{i = 1}^2 q_i \Bigg[ H (p_i; p_j, n_j) + k^\mu \left( \frac{\partial}
  {\partial k^\mu} H (p_i - k; p_j, n_j) \right)_{k \rightarrow 0} \Bigg] 
  \prod_j J (p_j,n_j) \, , \nonumber 
\eeqa
where we have Taylor expanded up to next-to-soft order in $k$ in the second line. We 
may also rewrite the derivative in the second line as
\beq
  \left. \frac{\partial}{\partial k^\mu} \, H (p_i - k; p_j, n_j) \right|_{k \rightarrow 0}
  \, =  \, - \, \frac{\partial}{\partial p_i^\mu} \, H(p_i; p_j, n_j) \, . 
\label{changeder}
\eeq
The zeroth order term in \eq{AHexpand} now can be seen to vanish due to charge 
conservation (or rather color conservation in the QCD case), and the Ward identity of 
\eq{Ward} implies
\beq
  {\cal A}^H_\mu (p_i, k) \, = \,  \sum_{i = 1}^2 q_i \left( \frac{\partial}
  {\partial p_i^\mu} H (p_i; p_j, n_j) \right) \prod_{j = 1}^2  J(p_j,n_j) \, .
\label{AHexpand2}
\eeq
The reader may wonder at this point whether it might be possible to add a 
separate transverse contribution to the right-hand side. Such a contribution 
is argued to be absent at the level of the amplitude in Ref.~\cite{Low:1958sn} 
(see also~\cite{Bern:2014vva,Broedel:2014fsa}), based on gauge invariance
and locality considerations. Here we notice that the $n_i^\mu$ dependence
of individual factors in \eq{AHexpand2} may in principle allow for transverse 
${\cal O}(k^0)$ contributions, for example of the form $k^\mu/(n_i \cdot k)$. 
All such contributions must however cancel in the complete amplitude, and 
we will not include them in our analysis.

Returning now to the amplitude for emission from the jet functions, it is 
convenient, following Refs.~\cite{Grammer:1973db,DelDuca:1990gz}, to 
introduce, for each jet, a decomposition of the polarization sum\footnote{Our 
definition of the $G$ and $K$ tensors differs from that of Ref.~\cite{DelDuca:1990gz} 
due to the fact that we have taken the momentum $k$ to be outgoing. Note also
that in our calculation $k^2 = 0$, although most of the argument goes through also for
off-shell $k_\mu$.} as
\beq
  \eta^{\mu \nu} \, = \, G^{\mu \nu} + K^{\mu \nu} \, , \qquad
  K^{\mu \nu} (p; k) \, = \, \frac{(2 p - k)^\nu}{2 p \cdot k - k^2} \, k^\mu \, ,
\label{KGdef}
\eeq
so that $G_{\mu \nu}$ satisfies
\beq
  p^\mu G_{\mu \nu} \, = \, {\cal O}(k) \, , \qquad G_{\mu \nu} k^\nu \, = \, 0 \,.
\label{satG}
\eeq
This decomposition leads  to a particularly simple expression for the emission of a
$K$-gluon from the jets. Indeed, the Ward identity in \eq{WardJ}, together with 
\eq{AmuJ}, yields
\beqa
  {\cal A}_\nu^{J_i} \, K_i^{\nu \mu} & = & q_i \, \frac{(2 p_i - k)^\mu}{2 p_i 
  \cdot k - k^2} \, H(p_i - k; p_j, n_j) \prod_{j = 1}^2 J(p_j, n_j) \nonumber \\
  & = & q_i \left[ \frac{(2 p_i - k)^\mu}{2 p_i \cdot k - k^2}
  \, {\cal A} - \left( K_i^{\nu \mu} \frac{\partial}{\partial p_i^\nu} \, H(p_i; p_j, n_j) \right)
  \prod_{j = 1}^2 J(p_j, n_j) \right] \, , 
\label{KAJ}
\eeqa
where $K_i^{\nu \mu}$ is the $K$ tensor appropriate to the $i$-th jet, we have again 
Taylor expanded in $k$, and we have recognized the non-radiative amplitude ${\cal A}$ 
in the second line. Upon combining this result with the emission from the hard function, as
given in \eq{AHexpand2}, one obtains 
\beq
  {\cal A}^{H, \mu} + \sum_{i = 1}^2 {\cal A}^{J_i}_{\nu} K^{\nu \mu}_i \, = \, 
  \sum_{i = 1}^2 q_i \left[ \frac{(2 p_i - k)^\mu}{2 p_i \cdot k - k^2} \, {\cal A}
  + \left( G^{\nu\mu}_i \frac{\partial}{\partial p_i^\nu} \, H(p_i; p_j, n_j) \right) 
  \prod_j J(p_j, n_j) \right] \, .
\label{HJsum}
\eeq
Using \eq{factorised}, one may rewrite this in a more useful form, as
\beqa
\label{HJsum2}
  {\cal A}^{H, \mu} + \sum_{i = 1}^2 {\cal A}^{J_i}_{\nu} K^{\nu \mu}_i & = &
  \sum_{i = 1}^2 q_i \Bigg[ \left( \frac{(2 p_i - k)^\mu}{2 p_i \cdot k - k^2}
  + G^{\nu \mu}_i \, \frac{\partial}{\partial p_i^\nu} \right) {\cal A} \nonumber \\
  & & \hspace{1cm} - \, H(p_i; p_j, n_j) \, G^{\nu\mu}_i \, \frac{\partial} {\partial p_i^\nu}
  \prod_{j = 1}^2 J(p_j, n_j) \Bigg] \, ,
\eeqa
so that derivatives with respect to hard momenta act only on the full non-radiative
amplitude, or on the process-independent jet functions.

We must now, finally, consider the emission of a $G$-gluon from the jets, which
is given by
\beq
  \sum_{i = 1}^2 {\cal A}^{J_i}_\nu \, G^{\nu \mu}_i \, = \, \sum_{i = 1}^2 \,
  G^{\nu\mu}_i \, H (p_i - k; p_j, n_j) \, J_\nu (p_i, k, n_i) \, \prod_{j \neq i} J(p_j, n_j) \, .
\label{Gjet}
\eeq
As argued in Ref.~\cite{DelDuca:1990gz}, projection with the $G$ tensor ensures that 
this contribution starts at next-to-leading power in the soft expansion. One may therefore
Taylor expand the shifted hard function, keeping only the zeroth order term. Adding this to 
\eq{HJsum2}, the radiative amplitude finally becomes
\beqa
\label{NEfactor}
  {\cal A}^\mu (p_j, k)  & = & \sum_{i = 1}^2 \Bigg[ \, q_i \left( \frac{(2 p_i - k)^\mu}{2 p_i 
  \cdot k - k^2} + G^{\nu \mu}_i \, \frac{\partial}{\partial p_i^\nu} \right) {\cal A} (p_i; p_j)  \\
  & & \hspace{1mm} + \, {\cal H} (p_j, n_j) \, \overline{\cal S}(\beta_j, n_j) \, G^{\nu \mu}_i
  \left( J_\nu (p_i, k, n_i) - q_i \, \frac{\partial}{\partial p_i^\nu} J(p_i, n_i) \right)
  \prod_{j \neq i} J(p_j, n_j) \Bigg] \, , \nonumber
\eeqa
where we have restored the factorized expression for the non-collinear function $H$ in
terms of the hard factor ${\cal H}$ and the reduced soft factor $\overline{\cal S}$.

In the remainder of the paper, we will provide strong evidence that \eq{NEfactor},
first derived with slight modifications in Ref.~\cite{DelDuca:1990gz}, provides the 
missing ingredient for a complete reconstruction of NLP threshold logarithms in
electroweak annihilation cross sections, generalizing the classic leading-power
factorization formula given in \eq{factorised}. Some comments are in order.
\begin{itemize}
\item We expect, and we will confirm below, that when \eq{NEfactor} is used to construct
a NNLO annihilation cross section, by contracting it with the corresponding tree-level 
matrix element and integrating over phase space, it will correctly generate all threshold 
logarithms up to NLP. Phase space integration does not present difficulties, since
next-to-soft contributions to \eq{NEfactor} can be integrated with the well-known factorized
expression for the phase-space measure at leading power, while correction to phase
space are only needed for leading-power matrix elements, and were discussed in 
detail in Ref.~\cite{Laenen:2010uz}.
\item A new universal quantity arises at this level, the {\it radiative jet function}, defined 
originally in Ref.~\cite{DelDuca:1990gz}. In order to predict NLP logarithms in specific 
scattering processes, one must calculate this quantity, which is the subject of the following 
section. 
\item The only process-dependent contributions on the right-hand side of \eq{NEfactor} 
are the full non-radiative amplitude ${\cal A}$ and the hard function ${\cal H}$. 
\item At the level of singular contributions, our result complements the analysis of 
Ref.~\cite{DelDuca:1990gz} by correctly accounting for the soft and soft-collinear 
regimes, and in particular by subtracting the double counting of the latter with the
inclusion of eikonal jet functions. 
\item It is important to note that our treatment suffices for
  electroweak annihilation cross sections, which do not involve
  final-state QCD jets, since all radiated gluons must be (next-to-)
  soft in this case. In the presence of final-state hadrons (for
  example in the case of Deep Inelastic Scattering (DIS), NLP
  threshold logarithms may in principle be associated with hard
  collinear emission. These are potentially not taken into account in
  \eq{NEfactor}, which relies upon the soft expansion.
\end{itemize}
Having now presented our general framework, we turn to the calculation of the radiative
jet function defined by \eq{Jmudef}, focusing in the present case on the abelian-like
contributions generated by the current in \eq{abcurr}, up to one loop.

%%%%%%%%%%%%%%%%%%%%%%%%%%%%%%%%%%%%%%%%%%%%%%

\section{The radiative jet function}
\label{sec:jetemit}

\eq{NEfactor} expresses the radiative amplitude for electroweak annihilation
in terms of a process-dependent hard function, together with a number of universal 
quantities. The soft function, the non-radiative jet and its eikonal counterpart
are already defined at leading power in the soft expansion, and can in principle 
be taken over without changes. One also encounters, however, a new universal
spin-dependent quantity: the radiative jet function of \eq{Jmudef}. This quantity is 
defined for finite emitted energy $k$, and must be computed with NLP accuracy
in the soft expansion, while keeping control of all collinear singularities which 
might interfere with the soft emission: all these contributions are expected to give 
rise to threshold logarithms when contracted with the tree-level complex-conjugate
amplitude and integrated over the soft gluon phase space.

To substantiate our claims, in this section we will compute the radiative jet function
to one loop order, selecting all terms that will contribute abelian-like threshold
logarithms, proportional to $C_F^2$, to the NNLO cross-section. We will therefore 
choose the current in \eq{abcurr}, omit graphs containing three-gluon vertices,
and pick the appropriate color structure for vertex corrections which would contribute
terms proportional to $C_A C_F$, which we disregard.  Before turning to that 
calculation, we will however pause to briefly discuss and motivate our choice of 
reference vectors $n^\mu$, which are necessary ingredients for non-radiative, 
eikonal and radiative jets.

%%%%%%%%%%%%%%%%%%%%%%

\subsection{Jet functions for light-like $n^\mu$}
\label{sec:nulljet}

It is customary in the factorization literature to compute the jet
functions appearing in \eq{factorised} with a reference vector off the
light cone, $n^2 \neq 0$ (see, for example,
Ref.~\cite{Collins:1989bt}). This has several advantages, in
particular in the case of all-order proofs and for renormalization
group studies.  Most notably, keeping $n^\mu$ off the light cone
prevents the appearance of collinear divergences associated with the
$n^\mu$ Wilson line, which are spurious and should in general be
removed; furthermore, keeping $n$ generic allows to test perturbative
calculations by verifying their independence on $n$; finally,
space-like $n$ preserves certain analytic properties of the amplitude
which are useful for all-order analyses. The non-radiative jet
function $J$, and its eikonal counterpart ${\cal J}$, are well known
and they have been computed at one loop for arbitrary (non-null) $n$
for example in Ref~\cite{Sterman:1986aj,Dixon:2008gr}.

\begin{figure}
\begin{center}
\scalebox{0.8}{\includegraphics{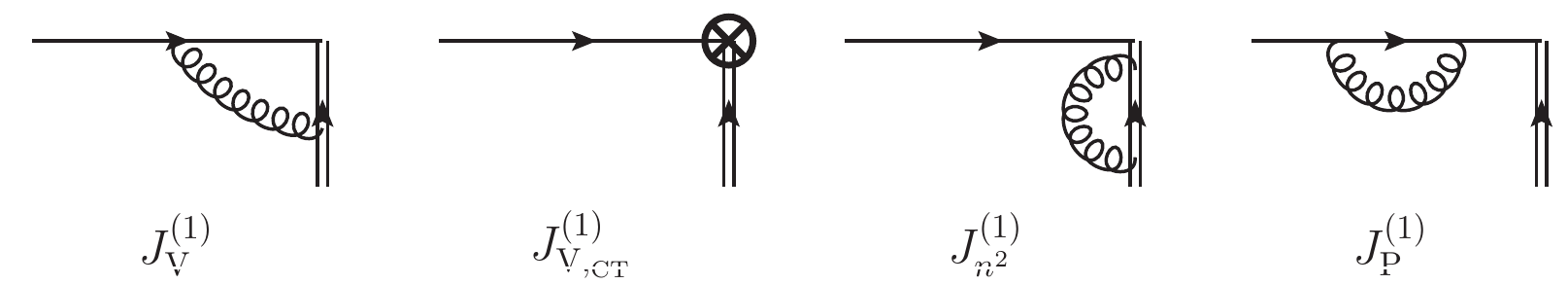}}
\caption{Feynman diagrams contributing to the one-loop jet function. Here 
$J_{V,_{CT}}^{(1)}$ denotes the counterterm associated with the vertex graph, 
$J_V^{(1)}$, while counterterms associated with external leg corrections have been 
omitted.}
\label{fig:jetdiags}
\end{center}
\end{figure}

Here we would like to argue that in the present case there are considerable 
computational advantages to be gained by keeping $n^2 = 0$, and at the same 
time the issues of interpretation, that can be quite subtle for generic correlators,
can easily be dealt with for the radiative amplitude that we are discussing here. 
To illustrate the problem, let us begin by considering the one-loop calculation of 
the non-radiative jet function $J(p,n)$. For an external incoming quark, one must 
consider the diagrams in Fig.~\ref{fig:jetdiags}, where the second graph denotes a 
UV counterterm associated with the renormalization of the vertex, and we have 
omitted the similar counterterm graphs associated with external leg corrections. 
For $n^2 \neq 0$, the complete jet function must depend on the dimensionless 
variable $z \equiv (p \cdot n)^2/(n^2 \mu^2)$, as a consequence of the invariance 
of the eikonal Feynman rules under the rescaling $n^\mu \to \kappa n^\mu$. 
Furthermore, the function is well defined in dimensional regularization due to the 
presence of the energy scale $p \cdot n$. For $n^2 = 0$, both of these properties 
are lost: dependence on $p \cdot n$ is in principle ruled out by rescaling invariance,
so that all integrals arising in the relevant Feynman diagrams are effectively 
scale-less, and must be defined to vanish in dimensional regularization. In fact,
there is an extra twist: while the diagrams vanish, one finds that there is a
residual dependence on $p \cdot n$ in the UV counterterms, due to an anomalous
breaking of rescaling invariance originating from the collinear pole associated
with emission from the light-like Wilson line~\cite{Dixon:2008gr,Gardi:2009qi}. 

To give a concrete example, consider the first diagram in Fig.~\ref{fig:jetdiags},
which we denote by $J_{\rm V}^{(1)}$. In $d = 4 - 2 \epsilon$ dimensions, and 
for $n^2 = 0$, it is given by
\beqa
  J_{\rm V}^{(1)} (p, n; \epsilon) & = & 2 {\rm i} \mu^{2 \epsilon} g_s^2 \int \frac{d^d k}{(2 \pi)^d}
  \frac{(\slsh{p} - \slsh{k}) \slsh{n}}{k^2 \, 2 n \cdot k \, (p - k)^2} \\
  & = & 2 {\rm i} \mu^{2 \epsilon} g_s^2 \int \frac{d^d k}{(2 \pi)^d} 
  \int_0^1 d x \int_0^1 d y \frac{2 y \, (\slsh{p} - \slsh{k}) \slsh{n}}{\big[ 
  y k^2 - 2 x y k \cdot p + 2 (1 - y) n \cdot k \big]^3} \, . \nonumber 
\label{Jadef}
\eeqa
In the second line, we have introduced Feynman parameters, and we note the 
characteristic parameter dependence of the $k^2$ term in the denominator,
which arises in the presence of linear denominators. Carrying out the momentum 
integration and using the Dirac equation, one may rewrite this as
\beq
  J_{\rm V}^{(1)} (p, n; \epsilon) \, = \, \frac{\alpha_s}{2 \pi} \, \left(4 \pi \mu^2 \right)^\epsilon \, 
  \Gamma(1 + \epsilon) \left( - 2 p \cdot n \right)^{- \epsilon} \, 
  \frac{1}{\epsilon (\epsilon-1)} \int_0^1dy \, y^{- 1 + \epsilon} \, 
  (1 - y)^{-1 - \epsilon} \, .
\label{Jacalc}
\eeq
At this point, one might be tempted to interpret directly the $y$ integral as 
$B(\epsilon, - \epsilon) = 0$. More accurately, one observes that the integral is
not well defined for any values of $\epsilon$, and must therefore be defined 
to vanish in dimensional regularization. In this simple case, it is actually easy
to disentangle the ultraviolet divergence (arising from the region $y \to 0$) from 
the infrared one (arising from the region $y \to 1$). One may simply insert a factor
of $(1 - y) + y = 1$ to see explicitly that infrared and ultraviolet poles cancel exactly;
using different regulators yields compatible results.
The standard treatment at this point is compute the {\it renormalized} jet function, 
isolating the ultraviolet divergence in \eq{Jacalc} and defining the appropriate 
$\overline{\rm MS}$ counterterm to subtract it. The sum of $J_V^{(1)}$ and 
$J_{V,_{CT}}^{(1)}$ is then simply the negative of the UV pole of $J_V^{(1)}$, 
which is correctly interpreted as an infrared divergence. For $n^2 = 0$ one 
recovers the expected double soft-collinear pole because of the explicit collinear 
divergence in \eq{Jacalc}. The external leg corrections in Fig.~\ref{fig:jetdiags}
can be treated similarly. In what follows, we will take the alternative route of
computing {\it bare} jet functions, which will allow us to simplify considerably all 
calculations, and make a more direct contact with the calculation performed
with the method of regions in Ref.~\cite{Bonocore:2014wua}, and eventually
with the SCET approach described in Ref.~\cite{Larkoski:2014bxa}. This approach
works for our current problem, and for the choice $n^2 = 0$, because of the
simple renormalization properties of \eq{NEfactor}. More specifically, we note 
that the left-hand side of \eq{NEfactor} is an on-shell scattering amplitude, and 
thus it is not renormalized: all counterterms  needed for the various factors must 
cancel. Furthermore, the right-hand side can be expressed almost entirely 
in terms of the complete non-radiative amplitude ${\cal A}$, which is also 
renormalization-group (RG) invariant. The first line of \eq{NEfactor} is already in 
this form; the second line can be brought to a similar form simply by noting
that, in each term in the sum over external partons, only one non-radiative
jet function is missing to reconstruct the full non-radiative amplitude. Multiplying
and dividing by that jet one can write \eq{NEfactor} as
\beqa
\label{NEfactor2}
  {\cal A}^\mu (p_j, k) & = & \sum_{i = 1}^2 \Bigg\{ \, q_i \left( \frac{(2 p_i - k)^\mu}{2 p_i 
  \cdot k - k^2} + G^{\nu \mu}_i \, \frac{\partial}{\partial p_i^\nu} \right) \\
  & & \hspace{1cm} + \, G^{\nu \mu}_i \left[ \frac{J_\nu (p_i, k, n_i)}{J(p_i, n_i)} - q_i \, 
  \frac{\partial}{\partial p_i^\nu} \Big( \ln J(p_i, n_i) \Big) \right]
  \Bigg\} \, {\cal A} (p_i; p_j) \, . \nonumber
\eeqa
In this form, it is evident that the factor in square brackets in the
second line must be RG invariant by itself, to all orders in perturbation theory.
This is easily verified by inspection, noting from the diagrammatic
expansion that the UV divergences of the radiative jet function
$J_\mu$ are the same as those of the non-radiative jet $J$, simply
multiplied by the tree-level expression for $J_\mu$. We are thus free
to compute \eq{NEfactor2} in either bare or renormalized perturbation
theory.  The advantage of using a light-like reference vector is now
apparent: with $n^2 = 0$, radiative corrections to the bare
non-radiative jet function vanish to all orders in perturbation
theory, and one can simplify \eq{NEfactor2} by setting
\beq  
  J (p_i, n_i) \, = \, 1 \, .
\label{J=1}
\eeq
The term containing the derivative of the non-radiative jet function now 
vanishes, and we are left with the simple expression~\footnote{Strictly 
speaking, one could write \eq{NEfactor3} explicitly in terms of the ratio
$\tilde{J}_\mu(p_i,k)=J_\mu(p_i,k)/J(p_i)$. As \eq{J=1} makes
clear, $J_\mu(p_i,k)=\tilde{J}_\mu(p_i,k)$ in dimensional
regularisation. As well as being UV finite, the ratio is also free of
soft (but not collinear) singularities due to the correspondence
between UV and IR poles. If a different regularisation scheme is used,
these properties persist for $\tilde{J}_\mu$, but not for $J_\mu$
itself.}
\beq
  {\cal A}^\mu (p_j, k) \, = \, \sum_{i = 1}^2 \left(q_i \, \frac{(2 p_i - k)^\mu}{2 p_i 
  \cdot k - k^2} + q_i \, G^{\nu\mu}_i \frac{\partial}{\partial p_i^\nu} + G^{\nu\mu}_i 
  J_\nu (p_i, k) \right) {\cal A} (p_i; p_j) \, .
\label{NEfactor3}
\eeq
We will use this explicit form in \secn{sec:DY}, and we believe that the simple formal
properties embodied in \eq{NEfactor2} and \eq{NEfactor3} will be useful also for the
all-order analysis that will have to be performed in order to construct a resummation
procedure for NLP threshold logarithms.

%%%%%%%%%%%%%%%%%%%%%%

\subsection{The radiative jet function at one loop}
\label{sec:jetemitcalc}

The main result of \secn{sec:nulljet} is that, for the purposes of the present calculation,
we are allowed to work with bare quantities, and with light-like reference vectors for jets. 
This has considerably simplified our task, since perturbative corrections to non-radiative
(eikonal) jet functions can be taken to vanish to all orders. The same is of course not true 
for the radiative jet function defined in \eq{Jmudef}, which in fact depends on several
momentum scales. We will now evaluate this function at one loop, using a light-like
reference vector, and working in bare perturbation theory as required by our reasoning
in \secn{sec:nulljet}.

We begin by defining the perturbative coefficients of the radiative jet function 
via\footnote{Given that we are explicitly talking about QCD corrections from now 
on, we have replaced the electromagnetic charge factor $q_i$ in section~\ref{sec:review}
with the strong coupling $g_s$. Note that in our convention the sign of $q_i$ is negative 
for an incoming quark line.}
\beq
  J_\nu \left( p, n, k \, ; \alpha_s, \epsilon \right)  \, = \, g_s \sum_{n=0}^{\infty}
  \left(\frac{\alpha_s}{4 \pi}\right)^n J^{(n)}_\nu \left(p, n, k \, ; \epsilon \right) \, .
\label{pertjmu}
\eeq
The operator definition in \eq{Jmudef} gives a straightforward result at tree level, 
consisting of a single emission from the external leg of momentum $p$, and yielding
the expression
\beqa
  J^{\nu(0)} \left(p, n, k \right) & = & \frac{\slash k \gamma^\nu}{2 p \cdot k}
  - \frac{p^\nu}{p \cdot k} \nonumber \\
  & = & - \, \frac{p^\nu}{p \cdot k} + \frac{k^\nu}{2 p \cdot k}
 - \frac{{\rm i} \, k_{\alpha} \Sigma^{\alpha \mu}}{2 p \cdot k} \, .
\label{Jnu0}
\eeqa
In the second line, we have chosen to decompose the result into spin-independent 
and spin-dependent parts, where the latter takes the form of a magnetic moment 
coupling to the fermion leg, involving the Lorentz generators
\beq
  \Sigma^{\alpha \nu} \, = \, \frac{\rm i}{4} \Big[ \gamma^\alpha, \gamma^\nu \Big] \, .
\label{sigmadef}
\eeq
The expression in the second line of \eq{Jnu0} naturally generalizes to the case
of emission from hard partons of different spin, simply by choosing the appropriate
form for the corresponding Lorentz generator.

\begin{figure}
\begin{center}
\scalebox{0.8}{\includegraphics{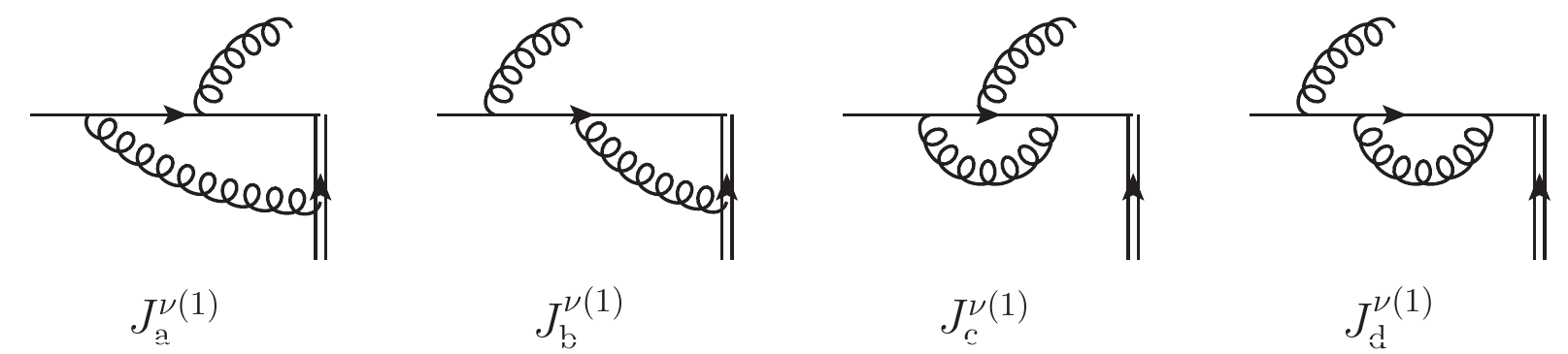}}
\caption{Contributions to the QED-like terms in the one-loop bare radiative jet function.}
\label{fig:Jmudiags}
\end{center}
\end{figure}

At one loop, things are more complicated, and the relevant contributions are depicted 
in Fig.~\ref{fig:Jmudiags}. Diagrams (a) and (b) have a more intricate kinematic 
dependence, as they link the physical external leg to the Wilson line, while diagrams 
(c) and (d) are standard radiative corrections to the emission amplitude. External leg
corrections are not included since they are pure counterterms.
The diagrams linking the parton line and the Wilson line are readily evaluated. For our 
purposes, we need to include terms up to ${\cal O} (\epsilon)$, since phase-space
integration over $k$ will generate singularities. On the other hand, we can neglect terms
quadratic in $k^\mu$ since they cannot contribute to NLP threshold logarithms. The
result is
\beqa
  J^{\nu(1)}_{a + b} \left(p, n, k \,; \epsilon \right) & = & \left( 2 p \cdot k \right)^{- \epsilon}
  \Bigg\{ \left( \frac{2}{\epsilon} + 4 + 8 \epsilon \right)
  \left[ \frac{n \cdot k}{p \cdot k} \frac{p^\nu}{p \cdot n}
  - \frac{\slsh{k} \gamma^\nu}{2 p \cdot k} - \frac{n^\nu}{p \cdot n} \right] 
  \nonumber \\ 
  & & + \, \left(1 + 3 \epsilon\right) \, \left( - \frac{2 k^\nu}{p \cdot k} + 
  \frac{\gamma^\nu \slsh{n}}{p \cdot n} - \frac{p^\nu \slsh{k} \slsh{n}}{p 
  \cdot k \, p \cdot n}\right) \Bigg\} \, .
\label{jmuab2}
\eeqa
Notice that this combination has only a single pole, while the individual diagrams (a) and (b)
have (canceling) double poles. Diagrams (c) and (d) do not depend on the reference vector 
$n$, and they yield
\beqa
  J^{\nu(1)}_{c + d} \left(p, n, k \,; \epsilon \right) & = & \left( 2 p \cdot k \right)^{- \epsilon}
  \Bigg[ \frac{1}{\epsilon} \left( \frac{\slsh{k} \gamma^\nu}{p \cdot k} + 
  \frac{k^\nu}{p \cdot k} \right) + \frac{5}{2} \, \frac{\slsh{k} \gamma^\nu}{p \cdot k} + 
  \frac{k^\nu}{p \cdot k} \\
  & & \hspace{3cm} + \, \, \epsilon \, \left( 5 \, \frac{\slsh{k} \gamma^\nu}{p \cdot k} + 
  2 \, \frac{k^\nu}{p \cdot k} \right) \Bigg] \, . \nonumber 
\label{jmucd2}
\eeqa
The full result for the bare one-loop radiative jet function is then given by
\beqa
  J^{\nu (1)} \left(p, n, k \,; \epsilon \right) & = & \left( 2 p \cdot k \right)^{- \epsilon} 
  \Bigg[ \left(\frac{2}{\epsilon} + 4 + 8 \epsilon \right) \left(\frac{n \cdot k}{p \cdot k}
  \frac{p^\nu}{p \cdot n} - \frac{n^\nu}{p \cdot n} \right)
  - (1 + 2 \epsilon) \, \frac{{\rm i} \, k_\alpha \Sigma^{\alpha \nu}}{p \cdot k} \nonumber \\
  & & \hspace{-2cm} + \, \left(\frac{1}{\epsilon}-\frac{1}{2}-3\epsilon\right)
  \frac{k^\nu}{p \cdot k} + \left(1 + 3 \epsilon \right) \left(\frac{\gamma^\nu \slsh{n}}{p \cdot n}
  - \frac{p^\nu \slsh{k} \slsh{n}}{p \cdot k \, p \cdot n} \right)
  \Bigg] + \mathcal O(\e^2, k) \, ,
\label{eq:nullj} 
\eeqa
where we have again chosen to write this in terms of spin-independent
and spin-dependent components, for the $n$-independent terms. A
comment is in order regarding the $\epsilon$ pole appearing in this
result. As explained in the previous section, we have avoided the
introduction of UV counterterms given that these will cancel when all
ingredients in eq.~(\ref{NEfactor3}) are combined. It follows that the
singularity appearing here can be interpreted directly as the infrared
pole that will appear in the final amplitude. 

An important check of the above results is the Ward identity, \eq{WardJ}. We note first 
that it is verified by the tree-level jet emission function in \eq{Jnu0}: indeed
\beq
  k_\nu \, J^{\nu(0)} \left(p, n, k \right) \, = \, - \, g_s J^{(0)} (p,n) \, ,
\label{Wardcheck1}
\eeq
where we have used the fact that, at tree-level,  $J^{(0)} (p, n) = 1$, and that our
expression has been derived for $k^2 = 0$, as well as our convention for the 
sign of the charge of an incoming fermion. At the one-loop level, the results of 
Eqs.~(\ref{eq:nullj}) and~\ref{Wardcheck1}) imply that
\beq
  k_\nu \, J^{\nu(1)} \, = \, 0 \, = \, - \, g_s J^{(1)} \, ,
\label{Wardcheck2}
\eeq
given that loop corrections to the non-radiative jet function vanish. Note again that 
we are verifying the Ward identity only up to corrections $\sim{\cal O}(k^2)$, since 
our jet functions have been computed for on-shell $k$.

In \eq{NEfactor2}, one must finally contract the jet emission function with the $G$ 
tensor defined in \eq{KGdef}. The reader may verify that, remarkably, the one-loop
expression in \eq{eq:nullj} is an eigenstate of $G$. More precisely one finds
\beq
  G^{\nu\mu} J^{(1)}_{\nu}  \left(p, n, k \right) \, = \, J^{(1)}_{\nu}  \left(p, n, k \right) \, .
\label{GJtot} 
\eeq
We now have everything we need in order to test our  factorized expression for
the radiative amplitude, \eq{NEfactor3}, in the context of a NNLO calculation. 
To this end, we consider Drell-Yan production in the following section.

%%%%%%%%%%%%%%%%%%%%%%%%%%%%%%%%%%%%%%%%%%%%%%

\section{Application to Drell-Yan production}
\label{sec:DY}

In the previous sections, we have assembled the ingredients needed to
organize threshold corrections at next-to-leading power, at the level
of scattering amplitudes.  As already mentioned above, we will now
test our results in the concrete example of Drell-Yan production at
NNLO, with the goal of reproducing abelian-like logarithms (weighted
by the color factor $C_F^2$) in graphs involving the real radiation
of a (next-to-) soft gluon, dressed by a virtual correction. This is a
non-trivial check that our formalism is correct, given that one
expects an interplay between soft and collinear contributions when
virtual gluons are included.

Drell-Yan production is the simplest testing ground for our formalism,
since it involves only two colored particles at leading order, it has
a non-trivial abelian-like limit, and it does not involve final state
jets; furthermore, NNLO corrections have been known for many
years~\cite{Hamberg:1990np,Hamberg2002403,Matsuura:1988sm,
Matsuura:1988nd,Harlander:2002wh}. A more interesting process for
phenomenological applications is Higgs production via gluon fusion,
where NLP logarithms are now known at
N$^3$LO~\cite{Anastasiou:2013mca,Herzog:2014wja} (see also
refs.~\cite{Li:2014bfa,Li:2014afw}): we plan to tackle that process in
future work, when the full non-abelian generalization of our formalism
has been worked out.

We recall that, for the purely real emission contributions, involving two real gluons at 
NNLO, all threshold logarithms at both LP and NLP have been shown to arise from the 
soft expansion, with no contamination from collinear singularities, as has been verified 
up to NNLO in Ref.~\cite{Laenen:2010uz}. Mixed real-virtual corrections at NNLO, which
we examine here, are the first instance in which the collinear generalization of Low's
theorem is required. In a diagrammatic framework, the various contributions involved
have been analysed recently in Ref.~\cite{Bonocore:2014wua}, using the method of 
regions of Refs.~\cite{Beneke:1997zp,Pak:2010pt,Jantzen:2011nz}. We will make
contact with this analysis in what follows.

%%%%%%%%%%%%%%%%%%%%%%

\subsection{The real-virtual Drell-Yan K-factor at NNLO}
\label{sec:DYreview}

We begin by very briefly setting up or notations and conventions. The leading-order 
Drell-Yan process for the production of an off-shell vector boson $V^*$ of invariant 
mass $Q^2$ proceeds through the process 
\beq
  q(p) + \bar{q} (\bar{p}) \rightarrow V^*(Q) \, ,
\label{DYLO}
\eeq
where $q$ and $\bar{q}$ denote a quark and antiquark respectively, and arguments 
label their four-momenta. Defining the squared centre-of-mass energy $s = (p + 
\bar{p})^2$, one may introduce the dimensionless threshold variable
\beq
  z \, = \, \frac{Q^2}{s} \, , 
\label{zdef}
\eeq
representing the fraction of available energy carried by the final state photon; the 
threshold limit then corresponds to $z \rightarrow 1$. The K-factor at fixed order in 
perturbation theory is defined by
\beq
  K^{(n)}(z) \, = \, \frac{1}{\sigma^{(0)}} \, \frac{d \sigma^{(n)}(z)}{dz} \, ,
\label{Kndef}
\eeq
where $\sigma^{(n)}$ is the $n$-loop Drell-Yan cross section. For the case of emission
of a single real gluon with momentum $k$, one may introduce the Mandelstam invariants
\beq
  t \, = \, - \, 2 k \cdot p \, , \qquad  u \, = \, - \, 2 k \cdot \bar{p} \, ,
\label{mandies}
\eeq
which may be parametrized as~\cite{Hamberg:1991np}
\beq
  t \, = \, - \, 2 s (1 - y)(1 - z) \, , \qquad u \, = \, - \, 2 s y (1 - z) \, ,
\label{tudef}
\eeq
where $0 < y < 1$. Using these variables, and setting, for simplicity, the renormalization scale 
$\mu^2 = Q^2$, the real-virtual contribution to the NNLO K-factor can be written as 
\beq
  K^{(2)}_{\rm r v} (z) \, = \, \frac{1}{16 \pi^2} \, \frac{(4\pi)^\epsilon}{\Gamma(2 - \epsilon)} \, 
  z^\epsilon \, (1 - z)^{1 - 2 \epsilon} \int_0^1 d y \, \big[ y (1 - y) \big]^{- \epsilon}
  \left[{\cal A}^\dag_{\rm r v} \, {\cal A}_{\rm r} + {\cal A}^\dag_{\rm r} \, {\cal A}_{\rm r v}
  \right] \, ,
\label{K1calc}
\eeq
where ${\cal A}_{\rm r v}$ and ${\cal A}_{\rm r}$ are the amplitudes for single real gluon 
emission, at NLO and LO respectively.

Although the complete NNLO Drell-Yan K-factor has been known for a long 
time~\cite{Hamberg:1991np}, no separate result exists in the literature for the 
real-virtual contribution. We have reproduced the relevant calculation, and we give 
here the result\footnote{Here and throughout, for brevity, we neglect terms involving 
transcendental constants,  as was done in Refs.~\cite{Laenen:2010uz,Bonocore:2014wua}. 
Such terms do not contain any new information.}  for threshold contributions up to NLP.  
One finds
\beqa
  K^{(2)}_{\rm r v} (z) & = & \left( \frac{\alpha_s}{4\pi} \, C_F \right)^2 \Bigg\{ \,
  \frac{32}{\epsilon^3} \, \Big[ {\cal D}_0 (z) - 1 \Big] +
  \frac{16}{\epsilon^2} \, \Big[ - 4 {\cal D}_1 (z) + 3 {\cal D}_0 (z) + 4 L (z) - 6 \Big] 
  \nonumber \\
  & & \hspace{21mm} + \, \frac{4}{\epsilon} \, \Big[ 16 {\cal D}_2 (z) - 
  24 {\cal D}_1 (z) + 32 {\cal D}_0 (z) - 16 L^2 (z) + 52 L (z)  - 49 \Big] \nonumber \\
  & & \hspace{21mm} - \, \frac{128}{3} {\cal D}_3 (z) + 96 {\cal D}_2 (z) - 256 {\cal D}_1 (z) + 
  256 {\cal D}_0 (z) \nonumber \\
  & & \hspace{21mm} + \, \frac{128}{3} L^3 (z) - 232 L^2 (z) + 412 L (z) - 408 
  \Bigg\} \, , 
\label{K1r1v}
\eeqa
where we defined 
\beq
  {\cal D}_n (z) \, = \, \left( \frac{\log^n (1 - z)}{(1 - z)} \right)_+ \, , \qquad 
  L(z) \, = \, \log(1 - z) \, .
\label{Didef}
\eeq
Further details of the calculation can be found in Appendix~\ref{app:DYcalc}.  Note that 
we do not include the Dirac $\delta$-function contributions in \eq{K1r1v}, as these mix
with virtual corrections that are irrelevant for the purposes of this paper.

%%%%%%%%%%%%%%%%%%%%%%

\subsection{Reconstructing the K-factor at NLP}
\label{KNLP}

We are now going to reproduce \eq{K1r1v}, starting from our factorized expression for 
the amplitude, given in \eq{NEfactor3}, and including the required contributions from
the phase space measure at NLP in the threshold expansion. Since the calculation is
not technically difficult, we simply outline the main steps and give the results for the
various contributions to the K-factor, distinguishing their physical origin.

%%%%%%%%%%%%%%%%%%%%%%

\subsubsection{NLP corrections to the phase-space measure}
\label{sec:phaspame}

Before discussing how the $\mathcal{D}_n (z)$ and $L^n (z)$ terms in
\eqref{K1r1v} are reproduced in our factorized approach, we address
the issue of how the amplitude factorization presented in
\eq{NEfactor} and \eq{NEfactor3} can be employed to obtain cross
section results\footnote{Note that the present situation differs from
  that of Ref.~\cite{Anastasiou:2013mca}, where the three-loop results
  for the form factors contributing to Higgs and Drell-Yan production
  could be used to infer the coefficients of the distributions
  $\mathcal{D}_n (z)$ at N$^3$LO for the corresponding cross sections;
  in fact, that procedure does not extend to the $L^n (z)$
  terms.}. \eq{K1calc} expresses the required contribution to the
K-factor in terms of the density \beqa
\label{lpnlp}
  {\cal P} & = & {\cal A}^\dag_{\rm r v} \, {\cal A}_{\rm r} + {\cal A}^\dag_{\rm r} \, 
  {\cal A}_{\rm r v} \nonumber \\
  & = & {\cal P}_{\rm LP} + {\cal P}_{\rm NLP} + \ldots \, ,
\eeqa
where the second line denotes the expansion in powers of $1 - z$ up to NLP. This must
be combined with the overall phase-space prefactor 
\beq
  z^\epsilon (1 - z)^{1 - 2 \epsilon} \, = \, (1 - z)^{1 - 2 \epsilon}
  \Big[1 + \epsilon \, (1 - z) + {\cal O} \left( (1 - z)^2 \right) \Big] \, ,
\label{phaspa}
\eeq
which arises from the real gluon phase space. 

Working at next-to-leading power in $(1 - z)$ has the advantage that corrections to the 
phase space measure and to the matrix element do not interfere. One may, in fact, write
the differential cross section schematically as
\beq
  d \sigma \, = \, d \Phi_{3, {\rm LP}} \left( {\cal P}_{\rm LP} + {\cal P}_{\rm NLP} \right) +
  d \Phi_{3, {\rm NLP}} {\cal P}_{\rm LP} \, ,
\label{sigNLP}
\eeq where $d \Phi_{3, {\rm (N)LP}}$ denotes the three-particle phase
space at the required order in the threshold expansion. In the present
context, we see that we can proceed by integrating the full squared
matrix element, given by \eq{NEfactor3} contracted with the tree-level
amplitude, with the eikonal expression for the phase-space measure; we
can then include phase-space corrections, where however only the
leading-power squared matrix element is needed. Note that NLP
corrections to the phase space measure affect the cross section to all
orders, starting with the tree-level emission of a single gluon, as
was already noted in Ref.~\cite{Laenen:2010uz} in order to reproduce
the known NLP logarithms in the one-loop Drell-Yan K-factor. Appendix
B of that paper also included a discussion of how to obtain the NE
contribution to a general multi-gluon phase space.  In what follows,
given the simplicity of the one-gluon-emission calculation, we will
not make explicit use of \eq{sigNLP}, which however would be useful in
order to construct all-order expressions.

%%%%%%%%%%%%%%%%%%%%%%

\subsubsection{From tree level to one loop}
\label{treetoone}

In order to reproduce all NLP terms in the real-virtual contribution to the NNLO 
Drell-Yan K-factor, we must calculate all contributions appearing in \eq{NEfactor2}, 
before integrating over the real-gluon phase space according to \eq{K1calc}. Before 
performing this calculation at one-loop level, it is worth pausing to briefly remark 
upon what happens at tree level. In that case, the derivatives in \eq{NEfactor2} and
(\ref{NEfactor3}) act on momentum-independent leading-order quantities\footnote{Note
that the derivatives do {\it not} act on the momentum dependence implicit in the particle
wave functions, for example the quark spinors in this case.}, and thus vanish. The
factor including the radiative jet function is easily computed using \eq{KGdef} and 
\eq{Jnu0}, and it gives
\beq
  G^{\nu \mu}J^{(0)}_{\nu} (p,n,k) \, = \, - \, \frac{{\rm i} \, k_\alpha \, 
  \Sigma^{\alpha\mu}}{p \cdot k_2} J^{(0)} (p,n) \, ,
\label{GJ0}
\eeq
where we have used the fact that $J^{(0)} (p,n) = 1$. Using the leading-power
factorization formula in \eq{factorised}, one then writes
\beq
  {\cal H} (p_i, n_i) \, \bar{\cal S}(p_i, n_i) \, G^{\nu \mu}_i J_{\nu} (p_i, n_i, k) 
  \prod_{j \neq i} J(p_j, n_j) \, = \, - \, \frac{{\rm i} \, k_\alpha \, 
  \Sigma^{\alpha \mu}}{p_i \cdot k} \, {\cal A} (p_i, p_j) \, .
\label{GJ0b}
\eeq
This combines with the first term on the right-hand side of \eq{NEfactor3} to reproduce 
the non-radiative amplitude dressed by a complete (spin-dependent) gluon emission. 
Upon integrating over the real gluon phase space, this precisely reproduces the 
NLO K-factor calculation, already carried out using an effective approach in 
Ref.~\cite{Laenen:2010uz}.

Returning now to the one-loop example, it is useful to distinguish
three contributions to the cross section which are physically
distinct. Let us begin by rewriting the one-loop contribution to
\eq{NEfactor3} as \beqa
\label{NEfactor4}
  {\cal A}^{\mu, (1)} (p_j, k) & = & \sum_{i = 1}^2 \Bigg[ \left( \frac{(2 p_i - k)^\mu}{2 p_i 
  \cdot k - k^2} + G^{\nu \mu}_i \frac{\partial}{\partial p_i^\nu} + G^{\nu \mu}_i 
  J^{(0)}_\nu (p_i, k) \right) {\cal A}^{(1)} (p_i; p_j) \nonumber \\ 
  & & \hspace{2cm} + \, G^{\nu \mu}_i 
  J_\nu^{(1)} (p_i, k) {\cal A}^{(0)}(p_i; p_j) \Bigg] \, ,
\eeqa
where we expanded the amplitudes in powers of $\alpha_s/(4 \pi)$, as in \eq{pertjmu}.
Using \eq{GJ0}, one finds that the first and third terms on the right-hand side combine to 
give the non-radiative amplitude, dressed by a complete (spin-dependent) gluon emission. 
This is directly analogous to the tree-level behaviour described above. Together with the 
remaining terms, there are then three contributions to calculate:  the dressed non-radiative
amplitude, the derivative of the non-radiative amplitude, and the radiative jet contribution. 
We compute them in turn.

%%%%%%%%%%%

\subsubsection{The dressed non-radiative amplitude}
\label{nrapse}

The one-loop non-radiative amplitude can be easily obtained from the well-known result
for the one-loop quark form factor (see, for example, Ref.~\cite{Dixon:2008gr}). Setting 
$\mu^2 = Q^2$ and reabsorbing the factors of $\ln(4 \pi)$ and $\gamma_E$ in the
renormalization scale, as appropriate to the $\overline{\rm MS}$ scheme, it is given by 
\beq
  {\cal A}^{(1)} (z) \, = \, - \, \frac{\alpha_s}{4\pi} \, z^\epsilon \, \frac{\Gamma^2 (1 - \epsilon) 
  \Gamma(1 + \epsilon)}{\Gamma(1 - 2 \epsilon)} \left( \frac{2}{\epsilon^2} + 
  \frac{3}{\epsilon} + 8 + 16 \epsilon + {\cal O}(\epsilon^2) \right) \, ,
\label{formfactor}
\eeq
where the factor $z^\epsilon$ arises since the non-radiative amplitude must be computed
with $(p + \bar{p})^2 = s$.
As is implicit in the above discussion, this dresses the tree-level amplitude involving 
one real gluon emission. The relevant amplitudes were reviewed recently in 
Ref.~\cite{Laenen:2010uz}. The results for the squared matrix elements at leading 
and next-to-leading power are
\beq
  \left| {\cal A}_{\rm r, LP} \right|^2 \, = \, 16 \, (1 - \epsilon) \, g_s^2 \, \frac{s^2}{u t} \, , 
\label{onerealE}
\eeq
and
\beq
  {\cal A}_{\rm r, NLP}^\dag \, {\cal A}_{\rm r, LP} + {\cal A}_{\rm r, LP}^\dag \,
  {\cal A}_{\rm r, NLP}  \, = \, 8 \, (1 - \epsilon) \, g_s^2 \left( \frac{s}{t} + \frac{s}{u} \right) \, ,
\label{onerealNE}
\eeq
respectively, where again LP and NLP refer to the order in the expansion in powers 
of $(1 - z)$. To obtain the appropriate contribution to the Drell-Yan K factor at NNLO, 
one must now multiply the sum of Eqs.~(\ref{onerealE}, \ref{onerealNE}) by the form 
factor contribution in \eq{formfactor}. After carrying out the phase space integration for 
the emitted gluon according to \eq{K1calc}, the result is
\beqa
\label{K2FF}
  K^{(2)}_{\rm d B} (z) & = & \left( \frac{\alpha_s}{4\pi} C_F \right)^2 \Bigg\{
  \frac{32}{\epsilon^3} \, \Big[ {\cal D}_0 (z) - 1 \Big] +
  \frac{8}{\epsilon^2} \, \Big[ - 8 {\cal D}_1 (z) + 6 {\cal D}_0 (z) + 8 L(z) - 14 \Big] 
  \nonumber \\
  & & \hspace{2cm} + \, \frac{16}{\epsilon} \, \Big[ 4 {\cal D}_2 (z) - 6 {\cal D}_1 (z) + 
  8 {\cal D}_0 (z) - 4 L^2 (z) + 14 L (z) - 14 \Big] \nonumber \\
  & & \hspace{2cm} - \, \frac{128}{3} {\cal D}_3 (z) + 96 {\cal D}_2 (z) - 256 {\cal D}_1 (z) + 
  256 {\cal D}_0 (z) \nonumber \\
  & & \hspace{2cm} + \, \frac{128}{3} L^3 (z) - 224 L^2 (z) + 448 L (z) - 512 \Bigg\} \, .
\eeqa
Not surprisingly, all plus distributions are correctly reproduced by the dressed Born
amplitude: the derivative and radiative jet contributions are expected to be strictly 
NLP, as we verify below.

%%%%%%%%%%%

\subsubsection{The derivative of the non-radiative amplitude}
\label{dernra}

The second term on the right-hand side of \eq{NEfactor3} consists of the non-radiative 
amplitude, differentiated in turn with respect to each external momentum. Considering 
for example the derivative with respect to $p$, \eq{formfactor} gives
\beq
  G^{\nu \mu} (p, k) \, \frac{\partial{\cal A}^{(1)}}{\partial p^\nu} \, = \, 
  \left[ - \frac{\epsilon}{p \cdot \bar{p}} \left(- p^\mu
  + \frac{\bar{p} \cdot k}{p \cdot k} \, \bar{p}^\mu \right) \right] {\cal A}^{(1)} \, .
\label{Gderivres}
\eeq
This derivative, and the corresponding one with respect to $\bar{p}$, must be contracted 
with the tree-level real-emission amplitude, before being integrated over phase space as 
in \eq{K1calc}. Including also complex conjugate graphs, one finds that the contribution 
to the K factor is
\beq 
  K^{(2)}_{\partial {\cal A}} (z) \, = \, \left(\frac{\alpha_s}{4\pi} \, C_F \right)^2 
  \Bigg\{ \frac{32}{\epsilon^2} + \frac{16}{\epsilon} \, \Big[ - 4 L(z)  + 3 \Big]
  + 64 L^2 (z) - 96 L (z) + 128 \Bigg\} \, ,
\label{Kderiv}
\eeq
which contributes at NLP, as expected.

%%%%%%%%%%%

\subsubsection{The radiative jet contribution}
\label{jetemcontr}

The last part of the NLP K-factor comes from the second line of
\eq{NEfactor}, and is due to those radiative jet contributions that
have not yet been included in the dressed non-radiative
amplitude. These can be extracted from \eq{eq:nullj}.  At this stage,
it becomes inevitable to make a precise choice for the factorization
vectors $n_i^\mu$ in the radiative jet functions. In the leading-power
factorization formula, \eq{factorised}, it is possible to engineer a
complete cancellation of the $n$ dependence of the non-radiative jets:
$n$-dependent poles cancel between jets and eikonal jets, while
$n$-dependent non-singular terms are cancelled by the hard function
${\cal H}$. At NLP, we have already made a special choice by setting
$n_i^2 = 0$; with this choice the presence of $n$-dependent poles in
\eq{eq:nullj} suggests that in order to achieve formal independence
from $n$ we would need to introduce a further subtraction, replacing
collinear poles in the direction $n_i$ with those associated with the
physical momentum of the parton colliding with parton $i$. There is
however a simple shortcut, already introduced in \eq{n1p2}: we may
simply choose \beq n \, = \, \bar{p} \, , \qquad \bar{n} \, = \, p \,
.
\label{npbar}
\eeq This is physically motivated by the fact that the Wilson line in
the direction $n_i$ acts as a replacement for the parton colliding
with parton $i$; furthermore, considering for example emission from
the parton with momentum $p$, one observes that $\bar{p}$ is the only
other light-like vector in the problem; finally, the method-of-regions
calculation of Ref.~\cite{Bonocore:2014wua} shows that this choice of
reference vectors does indeed account for all the poles in the
radiative amplitude. For processes with more legs, one would then
choose $n_i$ to be the anti-collinear vector appearing in the Sudakov
decomposition related to leg $i$, i.e. satisfying
Eq.~(\ref{nihatdef}), with $Q$ the appropriate hard scale.

With the choice in \eq{npbar}, we proceed by contracting \eq{eq:nullj} (for 
both the $p$ and $\bar{p}$ legs) with the NLO tree-level amplitude, including 
complex conjugate diagrams, and then we integrate over phase space. We find
\beq
  K_{\rm collinear}^{(2)} (z) \, = \, \left( \frac{\alpha_s}{4\pi} \,C_F \right)^2
  \Bigg\{ - \frac{16}{\epsilon^2} + \frac{4}{\epsilon} \, \Big[ 12 L(z) - 5 \Big] 
  - 72 L^2 (z) + 60 L(z) - 24 \Bigg\} \, ,
\label{Kcoll}
\eeq
where the label `collinear' for this contribution will be discussed in the following 
section. Once again, as expected, this contribution is strictly NLP.

%%%%%%%%%%%%%%%%%%%%%%

\subsection{Discussion}
\label{sec:discuss}

We have now calculated all necessary ingredients to reproduce the real-virtual,
abelian-like contribution to the NNLO Drell-Yan K factor. Adding these together
according to
\beq
  K^{(2)}_{\rm r v} (z) \, = \,  K^{(2)}_{\rm dB} (z) + K^{(2)}_{\partial{\cal A}} (z) + 
  K^{(2)}_{\rm collinear} (z) \, ,
\label{K2tot}
\eeq and using the results for each term given in Eqs.~(\ref{K2FF},
\ref{Kderiv}, \ref{Kcoll}), we precisely reproduce the full
perturbative result, given in \eq{K1r1v}. This is a non-trivial check
of the validity of the NLP factorization formula, \eq{NEfactor3},
particularly given the fact that all terms in the cross section
(including even the constant in the finite part) are reproduced: no
complete prediction for single-logarithmic contributions at NLP and
NNLO has been given before. 

At this point, it is interesting to compare our results with the
recent study of Ref.~\cite{Bonocore:2014wua}, which used the method of
regions~\cite{Beneke:1997zp, Pak:2010pt,Jantzen:2011nz} to classify
the same contributions to the NNLO Drell-Yan K-factor. The diagrams of
Fig.~\ref{sigbdiags} were calculated by expanding the virtual momentum
$k_1$ in soft, hard and (anti-)collinear regions (the latter with
respect to both legs $p$ and $\bar{p}$). This involves decomposing the
virtual momentum in a Sudakov decomposition, according to \beq k_1^\mu
\, = \, \frac{1}{2} \left( n_- \cdot k_1 \right) n_+^\mu + \frac{1}{2}
\left( n_+ \cdot k_1 \right) n_-^\mu + k_{1 \perp}^\mu \, ,
\label{Sudakov}
\eeq
where the vectors $n_\pm$ were chosen to be parallel to the $p$ and $\bar{p}$ 
directions, respectively. In fact, they correspond to the vectors $\hat{n}_i$ introduced 
here in \secn{sec:review}. It was then found that the only contributions to the K-factor 
originated from the hard and (anti-)collinear regions. The connection with the 
present study is as follows. The contributions from the hard region consist of the 
first two terms on the right-hand side of \eq{K2tot}, namely to the one-loop non-radiative 
amplitude dressed by an additional emission, and the derivative of the same amplitude. 
The remaining term in \eq{K2tot} comprises contributions from the collinear regions
associated with each leg. We see this directly in \eq{GJtot}, in which the terms on 
the right-hand side 
are explicitly proportional to an overall scale factor of $(2 p \cdot k)^{- \epsilon}$ or 
$(2 \bar{p} \cdot k)^{- \epsilon}$. This is the only scale that survives in the collinear
region, and it is precisely these contributions that the original Low's theorem~\cite{Low:1958sn} 
fails to capture, but which are discussed extensively in Ref.~\cite{DelDuca:1990gz} 
(see also Ref.~\cite{Larkoski:2014bxa}). Note that the exact correspondence is achieved 
because we have chosen here the $n_i$ vectors in accordance with a method-of-regions, 
or effective-theory approach. 

It is also interesting to examine our results in light of the recently proposed next-to-soft 
theorems in gravity~\cite{Cachazo:2014fwa} and gauge theories~\cite{Casali:2014xpa}, 
discussed further in Refs.~\cite{Cachazo:2014fwa,Casali:2014xpa,Schwab:2014xua,
Bern:2014oka,He:2014bga,Larkoski:2014hta,Afkhami-Jeddi:2014fia,Adamo:2014yya,
Strominger:2013jfa,He:2014laa,Cachazo:2013hca,Cachazo:2013iea,Cachazo:2014dia,
He:2014cra,Zlotnikov:2014sva,Kalousios:2014uva,Du:2014eca,White:2014qia,Bianchi:2014gla}. 
First, one may note that the next-to-soft theorems as presented in Ref.~\cite{Cachazo:2014fwa,
Casali:2014xpa} involve an explicit coupling to the total angular momentum of each outgoing 
leg, a structure which is not immediately apparent in \eq{NEfactor}. This is ultimately due, 
however, to the fact that \eq{NEfactor} is presented in a form which is most straightforward 
for explicit calculation of the various contributions. Nevertheless, one does see this 
structure emerge at loop level\footnote{The tree-level case has already been examined in
Ref.~\cite{White:2014qia}.} in \eq{NEfactor3}, where the second and third terms on the 
right-hand side combine to give
\beq
  G^{\nu\mu}_i \left( \frac{\partial}{\partial p_i^\nu} + J_\nu^{(0)} \right)
  {\cal A}^{(1)} \, = \, \frac{{\rm i} \, k_2^\nu}{p \cdot k_2} \, \left[ L_{\mu\nu}^{(i)} + 
  \Sigma^{(i)}_{\mu \nu} \right] {\cal A}^{(1)} \, ,
\label{angmom}
\eeq
where $\Sigma^{(i)}_{\mu \nu}$ is the spin angular momentum associated
with leg $i$, defined in \eq{sigmadef}, and
\beq
  L_{\mu \nu}^{(i)} \, = \, x_{i \mu} p_{i \nu} - x_{i \nu} p_{i \mu} \, = \, {\rm i} \!
  \left( p_{i \mu} \, \frac{\partial}{\partial p_i^\nu} - p_{i \nu} \, \frac{\partial}{\partial 
  p_i^\mu} \right) 
\label{orbital}
\eeq
is the orbital angular momentum. \eq{angmom} is precisely the coupling to the total 
angular momentum observed in Ref.~\cite{Casali:2014xpa}. Note, however, that the 
final term in \eq{NEfactor3} involves a coupling to the spin angular momentum
only, and thus corresponds to an explicit breaking of the next-to-soft theorem at 
loop level. As discussed above, this breaking is associated with collinear effects.

A particular point of discussion in recent literature is whether or not the next-to-soft 
theorems of Refs.~\cite{Cachazo:2014fwa,Casali:2014xpa} receive corrections at
loop level. This has been related to the sequential order in which one performs 
the dimensional regularization and soft expansions, with~\cite{Cachazo:2014dia} 
advocating performing the soft expansion first. This was further discussed in
Refs.~\cite{Bianchi:2014gla,Bern:2014oka,Bern:2014vva}, with Ref.~\cite{Bern:2014oka} 
strongly arguing that the expansion in the dimensional regularization parameter $\epsilon$ 
should be carried out first. A concrete example was examined in Ref.~\cite{Bonocore:2014wua}, 
namely the real-virtual corrections to the Drell-Yan K factor, that we have also considered 
here. There, it was found that contributions with a logarithmic dependence on the 
radiated gluon momentum $k$ are generated in the amplitude from the collinear region. 
These are needed to reproduce the known K factor, and would be manifestly absent
upon carrying out the soft expansion before the $\epsilon$ expansion. In the present 
context, this corresponds to expanding in $k$ before carrying out the integration over 
the virtual gluon momentum $k_1$. The presence of such terms indicate that there 
are indeed loop corrections to next-to-soft theorems.

The results of this paper, building upon the earlier work of Ref.~\cite{DelDuca:1990gz},  
provide for the first time a description of this breaking in terms of a universal factorized 
expression. It is then interesting to trace these loop corrections in the NLP expression
for the amplitude, \eq{NEfactor}. As Ref.~\cite{Bonocore:2014wua} makes clear for 
the Drell-Yan example, the terms which are problematic from the point of view of the 
next-to-soft theorems, and which depend upon the ordering of the soft and $\epsilon$ 
expansions, are those with an overall power of $(p \cdot k)^{- \epsilon}$, or $(\bar{p}
\cdot k)^{- \epsilon}$. As already noted above, these stem from the collinear regions 
associated with the corresponding external legs, and appear here in the second line 
of \eq{NEfactor}, through the radiative jet function $J_\mu$. This is not surprising: the 
failure of the next-to-soft theorems at loop level means that extra information must
be inserted by hand in order to patch up the collinear regions. This explains the 
presence of the jet emission function, and its definition makes clear that such terms 
are universal.

%%%%%%%%%%%%%%%%%%%%%%%%%%%%%%%%%%%%%%%%%%%%%%

\section{Conclusion}
\label{sec:conclusion}

In this paper, we have considered the generalization of the soft-collinear factorization 
theorem for gauge-theory scattering amplitudes to include corrections which are 
responsible for next-to-leading power threshold logarithms in high-energy cross 
sections. We concentrated on electroweak annihilation cross sections, using the
Drell-Yan process as an example, since in this case final state radiation is forced 
to be (next-to-) soft near threshold, and the problem of NLP threshold logarithms can
in principle be attacked using the tools provided by the Low-Burnett-Kroll-Del Duca
theorem. The organization of these logarithms at finite order, and possibly to all
orders in perturbation theory, is likely to play a crucial role in improving the precision 
of collider physics predictions, and it might also have a number of more formal 
applications, concerning the infrared limit of scattering amplitudes beyond divergent
contributions.

Building upon the work of Ref.~\cite{DelDuca:1990gz}, we have
constructed an explicit expression for the radiative scattering
amplitude, in which a (next-to-) soft gluon is emitted into the final
state of the Drell-Yan process. Our basic result is best summarized in
\eq{NEfactor2}, which expresses NLP corrections to the amplitude in
terms of known universal quantities that appear in the soft-collinear
factorization formula, as well as a new universal quantity, the {\it
  radiative jet function}, defined in terms of an auxiliary
`factorization vector' $n$ in a manner similar to ordinary
(non-radiative) jets. This function first appeared in
Ref.~\cite{DelDuca:1990gz}, and is computed here for the first time at
the one-loop level. We have also refined the treatment of
Ref.~\cite{DelDuca:1990gz} by introducing the explicit factorization
of soft modes, including the correct treatment of the double counting
of soft-collinear singularities, and by studying the role played by
the factorization vectors $n_i$. In particular, we have noted that it
is possible, in this case, to work with light-like reference vectors,
$n_i^2 = 0$, without introducing any further double countings of
poles, or other inconsistencies, thanks to the simple renormalization
properties of the proposed NLP expression for the amplitude,
\eq{NEfactor2}. This has led us to a considerably simplified formula,
\eq{NEfactor3}, in terms of light-like reference vectors, which can be
directly applied to construct NLP-accurate cross sections.

In order to test our generalized factorization formula, we have used
it to reproduce the known NLP contributions in the abelian-like
real-virtual contribution to the Drell-Yan K-factor at NNLO. Purely
real emission (of two gluons) at this order was already understood for
this process in Ref.~\cite{Laenen:2010uz}, using results based only on
the soft expansion, pushed to next-to-soft level. It is only when
virtual gluons are present that one becomes sensitive to collinear
effects. Thus, our exercise is a highly non-trivial check of
\eq{NEfactor3}, in which all parts of the formula are tested.  We find
that we do indeed reproduce all NLP threshold contributions, including
even the constant term in the finite part, which is not
logarithmically enhanced.
This suggests that \eq{NEfactor} can be used even
at finite orders to obtain good approximate higher-order cross sections at NNLO and 
beyond. We emphasize that, while our current results were derived for the abelian 
color structure, using a QED-like expression for the current, the formalism can be 
naturally generalized to the full non-abelian theory, by using the non-abelian current 
in \eq{nabcurr}. When this is done, the process of Higgs production via gluon fusion
will become accessible to our treatment, as well as other gluon-initiated cross 
sections.

Our results are an essential first step to eventually develop a complete resummation 
formalism for NLP threshold logarithms, but, to this end, much work remains to be done. 
The next step will be to precisely work out and test the non-abelian version of the present
formalism, which will immediately lead to an application to Higgs production in the gluon
fusion channel. We will then need to consider in more detail the factorized expressions 
that arise at the level of cross sections, rather than amplitudes, including a fully systematic
treatment of multi-gluon phase space. Finally, a non-trivial step will be the generalization
to final state jets, which will require a slightly different treatment, due to the importance
of hard-collinear gluon contributions, which cannot be reached through the soft expansion.
Work on all these issues is in progress.

%%%%%%%%%%%%%%%%%%%%%%%%%%%%%%%%%%%%%%%%%%%%%%

\section*{Acknowledgments}

We thank Pietro Falgari, Gerben Stavenga and Marco Volponi for useful
discussions and collaboration during the early stages of this project;
we also thank Einan Gardi, Simon Caron-Huot and Duff Neill for more
recent discussions.  This work was supported by the Research Executive
Agency (REA) of the European Union under the Grant Agreements number
PITN-GA-2010-264564 (LHCPhenoNet) and PITN-GA-2012-316704
(HIGGSTOOLS); by MIUR (Italy), under contract 2010YJ2NYW$\_$006, and
by the University of Torino and the Compagnia di San Paolo under
contract ORTO11TPXK. DB and EL have been supported by the Netherlands
Foundation for Fundamental Research of Matter (FOM) programme 104,
entitled ``Theoretical Particle Physics in the Era of the LHC'', and
the National Organization for Scientific Research (NWO). CDW is
supported by the UK Science and Technology Facilities Council
(STFC). We are grateful to the Higgs Centre for Theoretical Physics at
the University of Edinburgh, where part of this work was carried out,
for warm hospitality. LM is especially grateful to Nikhef for extended
hospitality under the auspices of LHCPhenoNet, and EL to the Munich
Institute for Astro- and Particle Physics (MIAPP) of the DFG cluster
of excellence ``Origin and Structure of the Universe''.

%%%%%%%%%%%%%%%%%%%%%%%%%%%%%%%%%%%%%%%%%%%%%%

\appendix

%%%%%%%%%%%%%%%%%%%%%%%%%%%%%%%%%%%%%%%%%%%%%%

\section{The abelian-like real-virtual NNLO Drell-Yan K-factor}
\label{app:DYcalc}

\begin{figure}
\begin{center}
\scalebox{0.75}{\includegraphics{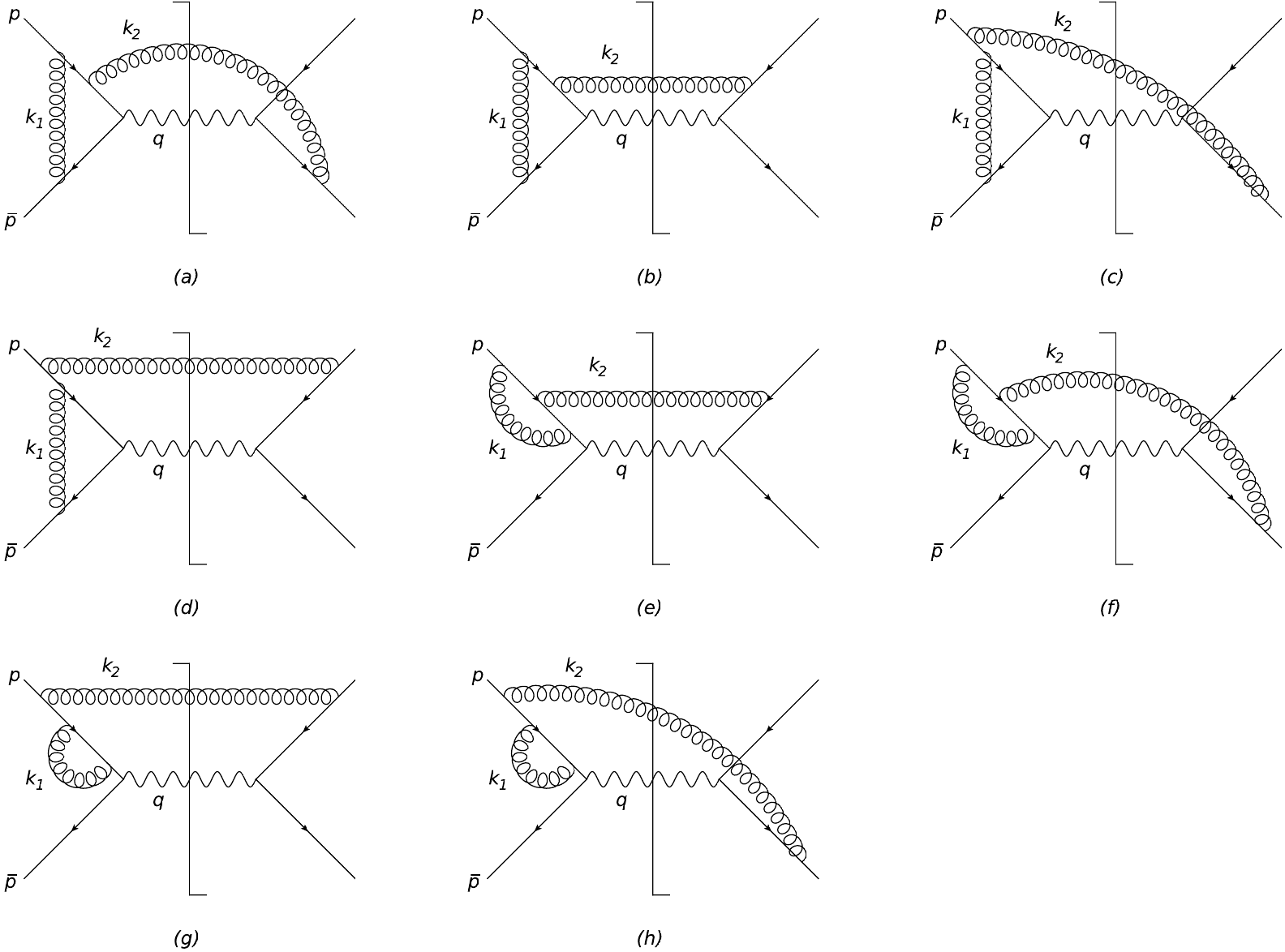}}
\caption{Feynman diagrams for the matrix element squared contributing to Drell-Yan
  production at NNLO, and involving one real and one virtual emission. Diagrams obtained 
  by interchanging $p\leftrightarrow \bar{p}$ and/or complex conjugation are not shown.}
\label{sigbdiags}
\end{center}
\end{figure}

In \eq{K1r1v}, we quoted a result for the $C_F^2$ part of the real-virtual contribution 
to the NNLO Drell-Yan $K$-factor. This has not previously appeared in the literature, so
we believe it is useful to collect a few details regarding the calculation. In particular, 
we show here, in Fig.~\ref{sigbdiags}, the relevant Feynman diagrams for the squared
matrix element, and we list below their individual contributions to the cross section at
NLP in the threshold expansion.
We have carried out the Dirac traces, the integration over the virtual momentum $k_1$, 
and the Passarino-Veltman reduction~\cite{Passarino:1978jh} using {\tt FeynCalc}~\cite{Kublbeck:1992mt}. 
Standard results for one-loop scalar integrals may be taken, for example, from 
Ref.~\cite{Ellis:2007qk}. After substituting the phase space parametrization in \eq{tudef}, 
one may expand the squared amplitude in powers of $(1 - z)$, where NLP accuracy 
corresponds to keeping the first subleading correction. One may then straightforwardly 
integrate over the real-gluon phase space, using \eq{K1calc}. Discarding transcendental 
constants, as in the text, the results for individual Feynman diagrams at NLP accuracy 
are given by
\beqa
\label{diags}
  (a): \quad & & \frac{24}{\epsilon^3} \, \Big[ {\cal D}_0 (z) - 1 \Big] + \frac{8}{\epsilon^2} \,
  \Big[ - 9 {\cal D}_1 (z) + 3 {\cal D}_0 (z) + 11 L (z) - 11 \Big] \nonumber \\ 
  & & + \, \frac{4}{\epsilon} \, \Big[ 27 {\cal D}_2 (z) - 18 {\cal D}_1 (z) + 12 {\cal D}_0 (z) - 
  37 L^2 (z) + 68 L (z) - 26 \Big] \nonumber \\
  & & - \, 108 {\cal D}_3 (z) + 108 {\cal D}_2 (z) - 144 {\cal D}_1 (z) + 
  96 {\cal D}_0 (z) \phantom{\Big[} \\
  & & + \, \frac{476}{3} L^3 (z) - 416 L^2 (z) + 352 L (z) - 208 \, , \nonumber \\ 
  (b): \quad & & \frac{8}{\epsilon^2} + \frac{8}{\epsilon} \, \Big[ - L(z) + 1 \Big] - 
  4 L^2 (z) + 32  \, , \\
  (c): \quad & & \frac{8}{\epsilon^3} \, \Big[ {\cal D}_0 (z) - 1 \Big] +
  \frac{8}{\epsilon^2} \, \Big[ {\cal D}_1 (z) + 3 {\cal D}_0 (z) - 3 L(z) - 1 \Big] \nonumber \\
  & & + \,  \frac{4}{\epsilon} \, \Big[ - 11 {\cal D}_2 (z) - 6 {\cal D}_1 (z) + 20 {\cal D}_0 (z) + 
  21 L^2 (z) - 20 L(z) - 20 \Big] \\ 
  & & + \frac{196}{3} {\cal D}_3 (z) - 12 {\cal D}_2 (z) - 112 {\cal D}_1 (z) + 160 {\cal D}_0 (z) 
  - 116 L^3 (z) + 224 L^2 (z) - 192 \, , \nonumber \\
  (d): \quad & & 0 \, , \\
  (e): \quad & & \frac{4}{\epsilon^2} + \frac{4}{\epsilon} \, \Big[ - 3 L(z) + 1 \Big] + 
  18 L^2 (z) - 12 L(z) + 8 \, , \\
  (f): \quad & & - \frac{12}{\epsilon^2} \, {\cal D}_0 (z) + \frac{12}{\epsilon} \, \Big[
  3 {\cal D}_1 (z) - {\cal D}_0 (z) + 1 \Big] \\ 
  & & - \, 54 {\cal D}_2 (z) + 36 {\cal D}_1 (z) - 24 {\cal D}_0 (z) - 36 L(z) \, , \nonumber  \\
  (g): \quad & & 0 \, , \\ 
  (h): \quad & & \frac{12}{\epsilon^2} \, \Big[ {\cal D}_0 (z)  - 1 \Big] + 
  \frac{12}{\epsilon} \, \Big[ - 3 {\cal D}_1 (z) + {\cal D}_0 (z) + 3 L(z) - 3 \Big] \nonumber \\ 
  & & + \, 54 {\cal D}_2 (z) - 36 {\cal D}_1 (z) + 24 {\cal D}_0 (z) - 
  54 L^2 (z) + 108 L (z) - 48 \, , 
\eeqa
where we have used the definitions of \eq{Didef}, and we have multiplied each diagram 
by four, in order to take into account complex conjugate diagrams, as well as those obtained 
from Fig.~\ref{sigbdiags} by interchanging $p$ and $\bar{p}$.

It is interesting to note that there are eikonal terms in graphs
(e)--(h), and they cancel when all such contributions are added
together. This had already been noted in Ref.~\cite{Bonocore:2014wua},
and the fact that such terms appear in individual diagrams is an
artifact of our (Feynman) gauge choice. Summing all diagrams together,
one obtains the result of \eq{K1r1v}.  Note also that we have not
included UV counterterms in the above calculation: if included, these
contributions cancel, in a manner reminiscent of the Ward identity
that prevents QCD corrections from renormalising the (QED)
photon-fermion vertex.

%%%%%%%%%%%%%%%%%%%%%%%%%%%%%%%%%%%%%%%%%%%%%%

\bibliography{NE_references}

%%%%%%%%%%%%%%%%%%%%%%%%%%%%%%%%%%%%%%%%%%%%%%

\end{document}